\documentclass{JHEP3}

\usepackage{epsfig}
\usepackage{amsbsy}
\usepackage{amssymb}
\usepackage{varioref}
\usepackage{pifont}

\def\sla#1{\ifmmode%
\setbox0=\hbox{$#1$}%
\setbox1=\hbox to\wd0{\hss$/$\hss}\else%
\setbox0=\hbox{#1}%
\setbox1=\hbox to\wd0{\hss/\hss}\fi%
#1\hskip-\wd0\box1 }
\newcommand{\half}{{\textstyle \frac{1}{2}}}
\newcommand{\fourth}{{\textstyle \frac{1}{4}}}

\title{Radiative Electroweak Symmetry Breaking \\ in a Little Higgs Model}

\author{Roshan Foadi, James T.~Laverty, Carl R.~Schmidt and Jiang-Hao Yu\\
Department of Physics and Astronomy, Michigan State University\\
East Lansing, MI 48824, USA\\
	E-mail: \email{foadiros@msu.edu}, \email{laverty1@msu.edu},
	\email{schmidt@pa.msu.edu}, \email{yujiangh@msu.edu}}

\abstract{
We present a new Little Higgs model, motivated by the deconstruction of a five-dimensional gauge-Higgs model.  
The approximate global symmetry is $SO(5)_0\times SO(5)_1$, breaking to $SO(5)$, with a gauged subgroup of $[SU(2)_{0L}\times U(1)_{0R}]\times O(4)_1$, breaking to $SU(2)_L \times U(1)_Y$.  Radiative corrections
 produce an additional small vacuum misalignment, breaking the electroweak symmetry down to $U(1)_{EM}$. 
Novel features of this model are: the only un-eaten pseudo-Goldstone boson in the effective theory is the Higgs boson; 
the model contains a custodial symmetry, which ensures that $\hat{T}=0$ at tree-level; and the potential for the Higgs boson is generated
entirely through one-loop radiative corrections.   A small negative mass-squared in the Higgs potential is obtained by a cancellation between
the contribution of two heavy partners of the top quark, which is readily achieved over much of the parameter space.  We can then obtain both a vacuum expectation value of $v=246$ GeV and a light Higgs boson mass, which is strongly correlated with the masses of the two heavy top quark partners.  For a scale of the global symmetry breaking of $f=1$ TeV and using a single cutoff for the fermion loops, the Higgs boson mass satisfies 120 GeV $\lesssim M_H\lesssim150$ GeV over much of the range of parameter space.  
For $f$ raised to 10 TeV, these values increase by about 40 GeV.  Effects at the ultraviolet cutoff scale may also raise the predicted values of the
Higgs boson mass, but the model still favors $M_H\lesssim 200$ GeV.

}

\preprint{{~MSUHEP--100104}}
\begin{document}

%%%%%%%%%%%%%%%%%%%%%%%%%%%%%%%%%%

\section{Introduction}
\label{sec:Intro}

The mechanism of electroweak symmetry breaking and the stabilization of the weak scale are two of the most important unresolved questions in particle physics. The Standard Model (SM) Higgs boson offers the simplest answer to the first question, but it leaves the second question unresolved.  In fact, the SM Higgs boson is unstable under quantum corrections, as its mass is naturally driven to the ultraviolet cutoff scale. Over the past decade a class of theories known as Little Higgs (LH) models has been proposed as a way to extend and stabilize the SM~\cite{Georgi:1974yw}--\cite{Thaler:2005en}. In LH models the Higgs boson is a pseudo-Goldstone boson of an approximate and spontaneously broken global symmetry. The latter is explicitly and {\em collectively} broken by extended gauge and Yukawa sectors, in such a way that the Higgs acquires a potential only if {\em two or more} couplings in the gauge or Yukawa sector are simultaneously switched on. Since quadratically divergent one-loop contributions to the Higgs mass can only arise from diagrams involving {\em one} coupling, it follows that these have to cancel. This is very similar to the supersymmetric scenario, in which the superpartners cancel the SM quadratic divergences. However in LH models the cancellation occurs between particles with the {\em same} spin, with interesting and extensively-studied collider signatures~\cite{Han:2003wu}--\cite{Belyaev:2006jh}.

Clearly, for a LH model to be realistic the generated Higgs potential must have a nonzero vacuum expectation value (vev). Furthermore, the electroweak vev $v$ must be much smaller than the vev $f$ associated with the spontaneous breaking of the larger symmetry group, since the main goal of any LH model is to naturally generate a hierarchy of scales between $v$ and the new-physics scale $f$.  This implies that the
ratio of the negative mass-squared, $m^2$, to the quartic coupling, $\lambda$, in the Higgs potential must be small in magnitude compared to $f^2$.
Typically in LH models, $m^2$ receives its dominant contribution from loops with the heavy partner of the top quark (which is required in the theory to cancel the quadratic divergence from the top-quark loop).  However, the dominant contribution to $\lambda$ is also typically generated
by loops of the same heavy top quark partner, so that a sufficiently large $\lambda$ is not generated radiatively.  For this reason, other
effective operators are introduced into the theory, whose coefficients depend on the details of the ultraviolet completion, but whose size can be estimated by naive dimensional analysis.   For instance, in the Moose-type models, such as the Minimal Moose~\cite{Arkani-Hamed:2002qx}, the quartic coupling arises from plaquette operators; in the Littlest Higgs~\cite{Arkani-Hamed:2002qy} the quartic coupling arises from a hard mass-squared for the additional scalars in the theory, which are then integrated out by equations of motion; and in the Simplest Little Higgs~\cite{simplest} model it arises from a small mass term for the scalars.   One disadvantage of this approach is that the unspecified coefficient of the new operator introduces an additional degree of unpredictability in the effective theory.  Furthermore, even with
the new contribution to $\lambda$, there must still be some amount of cancellation of the contribution to $m^2$ of the heavy top quark partner if one is to obtain a reasonably light Higgs boson~\cite{Casas:2005ev}.

A second requirement for the Higgs sector is the absence of large isospin violation. This is usually achieved by enlarging the overall global symmetry group to include $SU(2)_L\times SU(2)_R$, which in a LH model can be done minimally by imposing an $SO(5)$ symmetry~\cite{Chang:2003un}. This can create some problems in models with two Higgs doublets, with a potential which requires their vev's to be misaligned. This misalignment is a source of custodial isospin violation, which shows up in the form of dimension-six operators when the heavy states are integrated out. In Ref.~\cite{Chang:2003zn} this problem is avoided by constructing a model with a single Higgs doublet and an approximate custodial $SU(2)_C$, an extension of the Littlest Higgs with a coset $SO(9)/SO(5)\times SO(4)$. The electroweak constraints can also be weakened by introducing ``T-parity'', a new discrete symmetry under which the heavy fields are odd and the SM fields are even~\cite{Cheng:2003ju,Cheng:2004yc,Low:2004xc}. Then no effective operators are generated from tree-level exchanges of a single heavy field, since a vertex must contain an even number of these.

In this paper we introduce a LH model in which the only un-eaten scalar field is the Higgs boson, electroweak symmetry breaking is fully radiative, and an approximate custodial symmetry suppresses the sources of nonstandard isospin violation. The model is based on an $SO(5)_0\times SO(5)_1$ global symmetry, of which the $[SU(2)_{0L}\times U(1)_{0R}]\times O(4)_1$ subgroup is gauged.  The global and gauged symmetry structure is similar to
that of the Custodial Minimal Moose model~\cite{Chang:2003un}; however, in our model there is only one non-linear sigma field, with the result
that the Higgs boson is the only spin-zero field in the theory and there are no plaquette operators.  
The gauge sector of this model has also been considered in Ref.~\cite{Barbieri:2007bh}.  Our model is inspired from the deconstruction of an
$SO(5)\times U(1)_X$ gauge-Higgs model~\cite{Medina:2007hz}, which uses the fact that the $SO(5)$ structure is the minimal way to accommodate a gauge-Higgs and custodial symmetry.  In addition, it suggests the inclusion of fermions in terms of $SO(5)$
multiplets, with a simple implementation of  the LH mechanism in the Yukawa sector.  The novel feature of this fermion sector is that a second heavy top quark
partner produces canceling contributions to the $m^2$ term in the Coleman-Weinberg potential, so that it can easily be made small and negative.  As a consequence, the radiative Higgs quartic coupling, although small, is large enough to trigger spontaneous symmetry breaking with $v\ll f$, and the effective theory is more predictive than in LH models in which the quartic coupling arises from additional
operators.  In particular, the Higgs boson is naturally light in this model, with a mass that depends predominantly on a single mixing angle, $\sin^2\theta_t$, in the top quark sector.  For $f=1$ TeV and 10 TeV, we find $M_H\lesssim150$ GeV and $M_H\lesssim190$ GeV, respectively, over most of the range of $\sin^2\theta_t$.  Even after including effects of  unknown fermion dynamics at the cutoff scale, the assumption
that the Higgs potential is dominated by calculable contributions at one loop leads to a light Higgs boson over much of the parameter space.

The remainder of this paper is organized as follows. The gauge and fermion sectors of the theory are introduced in Sec.~\ref{sec:gauge} and \ref{sec:fermion}, respectively. In Sec.~\ref{sec:potential} we compute the Coleman-Weinberg potential and analyze the parameter space in which we can obtain both $v=246$ GeV and a light Higgs boson mass. In Sec.~\ref{sec:constraints} we compute the tree-level electroweak parameters, and derive the experimental bounds on the $SO(4)_1$ coupling ($g_1$) and $f$. Finally in Sec.~\ref{sec:conclusions} we offer our conclusions. Detailed calculations for the mass matrices and the Higgs potential can be found in the appendices.

\section{Gauge Sector}
\label{sec:gauge}

\EPSFIGURE[t]{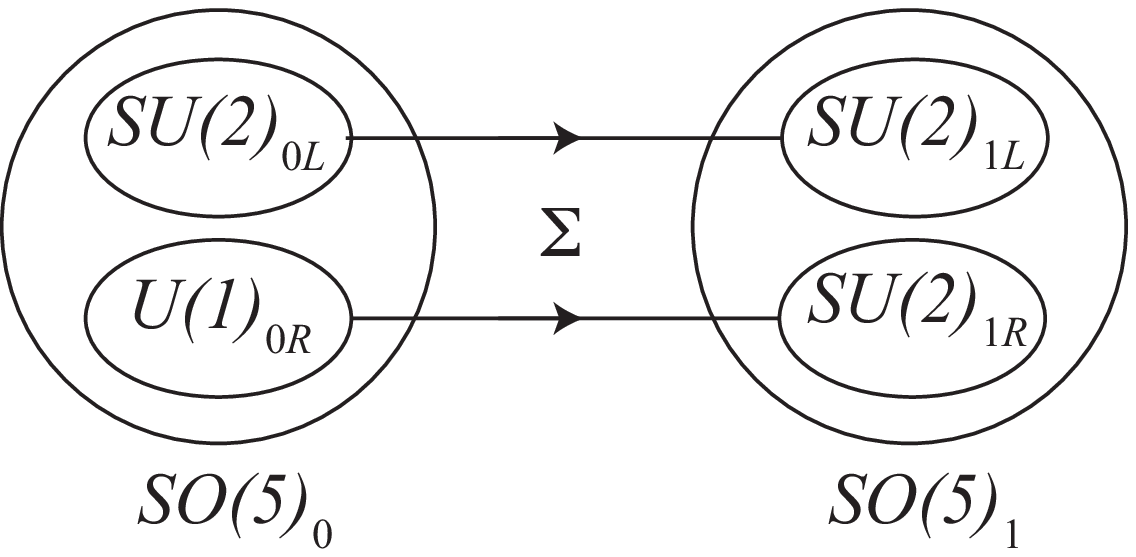,width=0.6\textwidth}
{Moose diagram for the model.  The approximate global symmetry is
$SO(5)_0\times SO(5)_1$, with an embedded gauge symmetry of
$\left[SU(2)_{0L}\times U(1)_{0R}\right]\times O(4)_{1}\cong
\left[SU(2)_{0L}\times U(1)_{0R}\right]\times\left[SU(2)_{1L}\times SU(2)_{1R}\times P_{1LR}\right]$.
\label{fig:moose}}

The gauge symmetry of our model is $SU(2)^3\times U(1)$, which is embedded in an approximate
$SO(5)\times SO(5)$ global symmetry.  The global symmetry is then broken spontaneously to the diagonal $SO(5)$ by a
non-linear sigma field.  This symmetry structure is represented in Fig.~\ref{fig:moose} by a moose diagram consisting of two sites, $0$ and $1$,
where the outer circles are the global $SO(5)$'s and the inner ellipses are the gauged subgroups.  In terms of the moose site indices, 
the global symmetry can be written $SO(5)_0\times SO(5)_1$, while the gauged subgroup is $\left[SU(2)_{0L}\times U(1)_{0R}\right]\times\left[SU(2)_{1L}\times SU(2)_{1R}\right]$.  In this description the $L$ and $R$ subscripts 
indicate the two commuting $SU(2)$ subgroups of $SO(5)$, while $U(1)_{0R}$ is a $U(1)$ subgroup of $SU(2)_{0R}$.
Note that the model can be considered a severe deconstruction of the 5-dimensional $SO(5)\times U(1)_X$ Gauge-Higgs model of Ref.~\cite{Medina:2007hz}, where the extra $U(1)_X$ 
symmetry has been removed.  In terms of this deconstruction, the sites $0$ and $1$ are the two end-branes of the 5-dimensional interval, while the non-linear sigma field plays the role of the fifth component of the gauge fields in the bulk.  

The non-linear sigma field is parametrized by
\begin{equation}
		\Sigma\ =\  e^{\sqrt{2}i\pi^A T^A/f}	\,,
\end{equation}
where we have chosen the normalization, ${\rm tr}\left(T^{A}T^{B}\right)=\delta^{AB}$,
so that the gauged $SU(2)$ sub-matrices have the conventional normalization.   
A convenient basis for the ten $SO(5)$ generator matrices is $\{T^{a}_L, T^{a}_R, T^1,T^2,T^3,T^4\}$, given in Appendix~\ref{sec:matrices} in Eq.~(\ref{genmatrices}).
Under an $SO(5)_0\times SO(5)_1$
transformation, the sigma field transforms as $\Sigma\rightarrow U_0\Sigma U^\dagger_1$, where $U_{0,1}$ are $SO(5)$ matrices in the fundamental representation.  Gauging the  $\left[SU(2)_{0L}\times U(1)_{0R}\right]\times\left[SU(2)_{1L}\times SU(2)_{1R}\right]$ subgroup leads to the following covariant derivative 
\begin{eqnarray}
		D^\mu \Sigma\  =\  \partial^\mu \Sigma     
				- i g_{{0L}} W_{{0L}}^{a\mu} T_L^a \Sigma 
				- i g_{{0R}} B_{{0R}}^{\mu} T_R^3 \Sigma
				+ i g_{{1L}} W_{{1L}}^{a\mu} \Sigma T_L^a 
				+ i g_{{1R}} W_{{1R}}^{a\mu} \Sigma T_R^a	\,.
				\label{covderivs}
\end{eqnarray}
With this we can write the Lagrangian density for the gauge and sigma fields as
\begin{eqnarray}
{\cal L}_{\rm gauge}&=& -{1\over4}W_{0L}^{a\,\mu\nu}W^a_{0L\,\mu\nu}
-{1\over4}B_{0R}^{\mu\nu}B_{0R\,\mu\nu}-{1\over4}W_{1L}^{a\,\mu\nu}W^a_{1L\,\mu\nu}
-{1\over4}W_{1R}^{a\,\mu\nu}W^a_{1R\,\mu\nu}\nonumber\\
&&
+{f^2\over4}{\rm tr}\Big[\left(D^\mu\Sigma\right)\left(D_\mu\Sigma\right)^\dagger\Big]\ .
\label{eq:lagrange}
\end{eqnarray}
In this paper we shall write $g_{1L}$ and $g_{1R}$ as if distinct. However, in models similar to ours it has been found that promoting an $SU(2)_L\times SU(2)_R$ gauge symmetry to $O(4)$ turns out to protect the tightly constrained $Z b_L \bar{b}_L$ coupling from large loop corrections~\cite{Agashe:2006at,Carena:2007ua,SekharChivukula:2009if}.  For this reason, and for simplicity, we will choose $g_{1L}= g_{1R}\equiv g_1$ for any computations, imposing the $L$-$R$ exchange symmetry $P_{1LR}$ necessary for the full $O(4)_1\sim SU(2)_{1L}\times SU(2)_{1R}\times P_{1LR}$. However, we will not compute the $Z b_L \bar{b}_L$ coupling, as well as other electroweak observables at one loop, leaving this for future work~\cite{flsy}.

With the gauged subgroups embedded in the global $SO(5)_0\times SO(5)_1$ as given by Eq.~(\ref{covderivs}), a vacuum alignment of $\langle\Sigma\rangle=1$ spontaneously breaks the gauge symmetry $\left[SU(2)_{0L}\times SU(2)_{1L}\right]\times\left[ U(1)_{0R}\times SU(2)_{1R}\right]$ down to the SM  electroweak group $SU(2)_L\times U(1)_{R=Y}$.
There are 6 exact Goldstone bosons, which will be eaten by 6 linear combinations of the gauge fields, giving them masses of order the symmetry breaking scale $f$.  The remaining 4 dynamical fields contained in $\Sigma$ have exactly the right quantum numbers to play the role of the standard model Higgs doublet $H$.  Although $H$ is not an exact Goldstone boson, we note that the gauge sector of the model has the collective symmetry breaking necessary to forbid any quadratic divergences to the Higgs effective potential at one loop. If we set the couplings to zero at either site 0 or at site 1, the global $SO(5)$ symmetry at that site becomes exact, and all 10 pion fields,
including the Higgs doublet, become exact Goldstone bosons.  Thus, any field-dependent term in the Higgs effective potential must have contributions collectively from both the couplings at site 0 and at site 1, which can only contain quadratic divergences at two loops or higher.

Working in unitary gauge, where we set the eaten Goldstone boson fields to zero, we can identify $H$ in $\Sigma$ by letting
\begin{eqnarray}
		\Pi \,\equiv\, \sqrt{2}\pi^A T^A\, =\,
					    \left(\begin{array}{cc}
					    	0_{4\times4} 
						&\left( 
						\begin{array}{c}
						H \\
						\tilde{H}\\
						\end{array}\right)\\
						\left(\begin{array}{cc}
						H^\dagger & \tilde{H}^\dagger\\
						\end{array}\right) & 0 \\	
		                             \end{array}\right)\,\ ,
\end{eqnarray}
where
\begin{equation}
H\,=\, \left(\begin{array}{c}
				h_1 \\
				h_2
		           \end{array}\right) \quad \mathrm{and} \quad 
\tilde{H}\, =\, -i \sigma_2 H^*\, = \,
			\left(\begin{array}{c}
				-h_2^* \\
				h_1^*
		           \end{array}\right)\,,
\end{equation}
with 
\begin{eqnarray}
				h_1 &=& \frac{1}{\sqrt{2}}(\pi^1 + i\pi^2 )\,,\\ \nonumber
				h_2 &=& \frac{1}{\sqrt{2}}(\pi^3 + i\pi^4 )\,.
\end{eqnarray}
Expanding and re-organizing the $\Sigma$ field, we obtain
\begin{eqnarray}
		\Sigma  &=&  e^{i\Pi/f}  \ =\ 1 +\frac{ i\Pi}{ \sqrt{2}|H|}  s
			  -\frac{\Pi^2}{2|H|^2}  
				 \left(1-c\right) \,,
				 \label{sigmaexpand}
\end{eqnarray}
where
\begin{eqnarray}
s&=&\sin\left(\frac{\sqrt{2}|H|}{f}\right)\qquad\mathrm{and}\qquad
c\, =\, \cos\left(\frac{\sqrt{2}|H|}{f}\right)\,,
\end{eqnarray}
and $|H|=(h_1^2+h_2^2)^{1/2}$.

Any further misalignment of the vacuum will result in a vacuum expectation value for the Higgs doublet, 
\begin{equation}
\langle H\rangle\,=\, \frac{1}{\sqrt{2}}\left(\begin{array}{c}
				v \\
				0
		           \end{array}\right)\,,
\end{equation}
breaking the gauge symmetry completely down to $U(1)_{EM}$.  Determining the value of $v$ requires an analysis
of the effective potential, which we do at one loop in this paper.  For this we need the mass terms for the gauge bosons, as
a function of the Higgs field, which we can take to be along the direction of its vacuum expectation value, without loss of generality.  Using the expression Eq.~(\ref{sigmaexpand}) for $\Sigma$ in the gauge Lagrangian, Eq.~(\ref{eq:lagrange}), we obtain
\begin{eqnarray}
		{\cal L}_{\rm mass} &=& \frac{f^2}{4}\Biggl\{ g_{0L}^2 W_{0L}^{a\mu} W_{0L\mu}^{a}
				+  g_{0R}^2 B_{0R}^{\mu} B_{0R\mu}
				+  g_{1L}^2 W_{1L}^{a\mu} W_{1L\mu}^{a}
				+ g_{1R}^2 W_{1R}^{a\mu} W_{1R\mu}^{a}\nonumber\\
			&&\qquad-2 (1-a) \, g_{0L} g_{1L} W_{0L}^{a\mu} W_{1L\mu}^{a}
			- 2a \, g_{0L} g_{1R} W_{0L}^{a\mu} W_{1R\mu}^{a}\nonumber\\
			&&\qquad- 2a \, g_{0R} g_{1L} B_{0R}^{\mu} W_{1L\mu}^{3}
			- 2(1-a)\, g_{0R} g_{1R} B_{0R}^{\mu} W_{1R\mu}^{3}\Biggr\}
	\,,\label{eq:gaugemass}
\end{eqnarray}
where
\begin{eqnarray}
		a\,=\, \frac{1}{2}\left(1-c\right)\, =\, \sin^2\left(\frac{|H|}{\sqrt{2}f}\right) \,.
\end{eqnarray}

For $a=0$ the mass matrices can be easily diagonalized.  The charged gauge boson masses are
\begin{eqnarray}
		M_{W^\pm}^2&=&0\nonumber\\
		M_{W_L^\pm}^2&=&\half\left(g_{0L}^2+g_{1L}^2\right)f^2\\
		M_{W_R^\pm}^2&=&\half g_{1R}^2f^2\,,\nonumber
\end{eqnarray}
and the neutral gauge boson masses are
\begin{eqnarray}
		M_{W^3}^2&=&0\nonumber\\
		M_{B}^2&=&0\nonumber\\
		M_{Z_L}^2&=&\half\left(g_{0L}^2+g_{1L}^2\right)f^2\\
		M_{Z_R}^2&=&\half \left(g_{0R}^2+g_{1R}^2\right)f^2\,.\nonumber
\end{eqnarray}
The massless states, $W^a$ and $B$, correspond to the unbroken $SU(2)_L\times U(1)_Y$ gauge symmetry.

\EPSFIGURE[t]{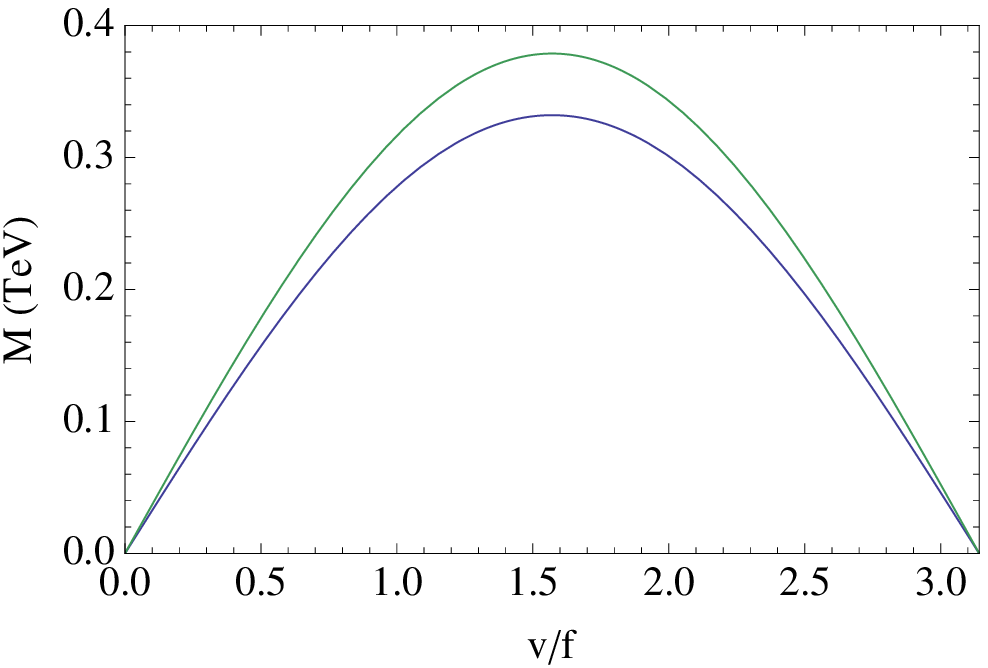,width=0.6\textwidth}
{Light gauge boson masses ($W$ and $Z$) as a function of $v/f$, for $g_1^2=6$ and $f=1$ TeV. The upper curve is $M_Z$ and the lower curve is $M_W$.  \label{fig:GmassL}}

\EPSFIGURE[t]{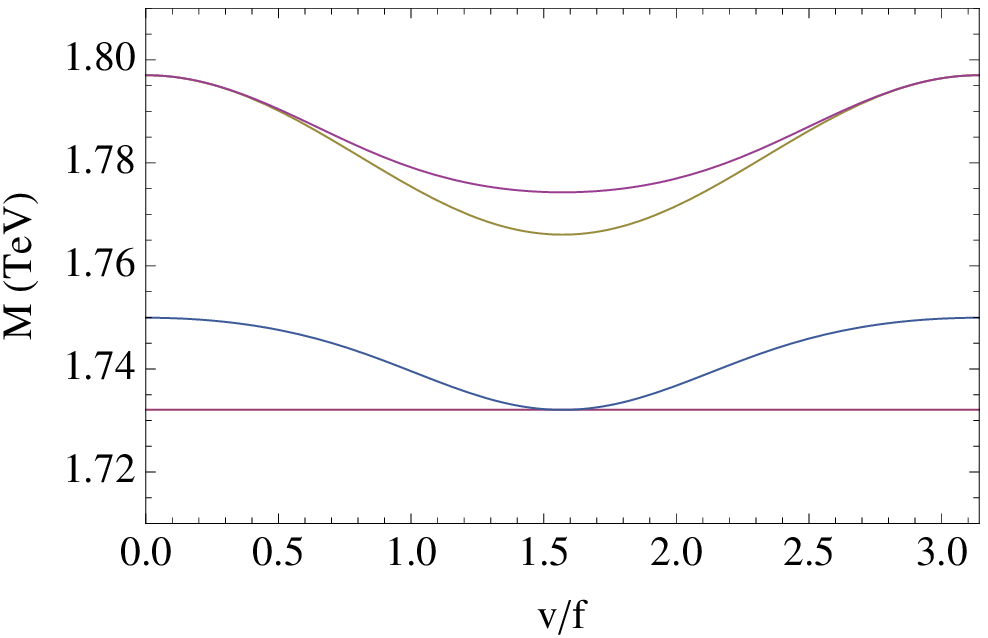,width=0.6\textwidth}
{Heavy gauge boson masses as a function of $v/f$, for $g_1^2=6$ and $f=1$ TeV.  The curves from top to bottom are $M_{Z_L}$, $M_{W_L}$, $M_{Z_R}$, and $M_{W_R}$.
\label{fig:GmassH}}

For a nonzero vacuum expectation value, $\langle|H|\rangle=v/\sqrt{2}$, it is also straightforward to solve
for the mass eigenvalues exactly.  There is one massless neutral boson, corresponding to the photon,
and the remaining neutral and charged gauge boson masses can be obtained as the solutions to two cubic 
characteristic equations.  
In Fig.~\ref{fig:GmassL} we plot the light $W$ and $Z$ boson masses and in Fig.~\ref{fig:GmassH} we plot the heavy gauge boson masses as a function of $v/f$ for
representative choices of the parameters: $g_1^2=6$ and $f=1$ TeV.  Clearly, for $f=1$ TeV the only allowed value of $v/f$ is $\sim$0.246, but it is nonetheless interesting to note the symmetry of the
solutions under the exchange of $(v/f)\leftrightarrow (\pi-v/f)$.  This is a result of the parity symmetry, $P_{1LR}$, which holds when
$g_{1L}=g_{1R}$.  Under this symmetry:
\begin{eqnarray}
		W_{1L}^{a\mu}\leftrightarrow W_{1R}^{a\mu}\nonumber\\
		\Sigma \rightarrow \Sigma^\prime\,=\,\Sigma P\,,\nonumber
\end{eqnarray}
with
\begin{eqnarray}
		P \  =\ 	    \left(\begin{array}{rrrrr}
						0 & 0 & 0 & -1 & 0 \\
						0 & 1 & 0 & 0 & 0 \\
						0 & 0 & 1 & 0 & 0 \\
						-1 & 0 & 0 & 0 & 0 \\
						0 & 0 & 0 & 0 & -1 \\	
		                             \end{array}\right)\ .
\end{eqnarray}
The matrix $P$ satisfies $PT^a_{L,R}P=T^a_{R,L}$.  It can be shown that the transformed field $\Sigma^\prime$ is
related to the original field $\Sigma$ by a shift of $v/f\rightarrow v/f+\pi$, up to an overall $O(4)_1$ transformation.
This, coupled with the discrete $H\leftrightarrow-H$ symmetry of the model, results in the symmetry of the mass solutions.

As required by a little Higgs model, we will want $v/f$ to be small. Thus, it is useful to solve for the masses and mixings 
perturbatively in $a\approx \left[v/\left(2f\right)\right]^2$. 
At leading nonzero order in $v/f$, the massless charged gauge bosons, $W^\pm$, gain a mass
\begin{eqnarray}
		M_{W^\pm}^2&\approx&\fourth g_{L}^2v^2\,,
\end{eqnarray}
while the massless neutral gauge bosons, $W^3$ and $B$, mix exactly as in the standard model to give the photon $A$ and the
$Z$ boson with masses
\begin{eqnarray}
		M_{A}^2&=&0\nonumber\\
		M_{Z}^2&\approx&\fourth\left(g_{L}^2+g_{R}^2\right)v^2\,,
\end{eqnarray}
where we have defined the couplings $g_L$ and $g_R$ by
\begin{eqnarray}
		\frac{1}{g_L^2}&=&\frac{1}{g_{0L}^2}+\frac{1}{g_{1L}^2}\nonumber\\
		\frac{1}{g_R^2}&=&\frac{1}{g_{0R}^2}+\frac{1}{g_{1R}^2}\,.\label{eq:gLgR}
\end{eqnarray}
Note that $g_L$ and $g_R$ play the roles of the standard model $SU(2)_L$ and $U(1)_Y$ gauge couplings, respectively.
Of course, the photon is exactly massless, being associated with the unbroken $U(1)_{EM}$, with coupling constant $e$ given
by
\begin{eqnarray}
		\frac{1}{e^2}&=&\frac{1}{g_{L}^2}+\frac{1}{g_{R}^2}\,=\,\frac{1}{g_{0L}^2}+\frac{1}{g_{1L}^2}+\frac{1}{g_{0R}^2}+\frac{1}{g_{1R}^2}\,.
\end{eqnarray}
More details of the gauge boson masses and mixings are given in Appendix~\ref{sec:gaugemass}.

\section{Fermion Sector}
\label{sec:fermion}

In this section, we will consider only one generation of quarks, although 
multiple generations of quarks and leptons can be incorporated as well.  We are motivated by the deconstruction of the 5-dimensional $SO(5)\times U(1)_X$ Gauge-Higgs model of Ref.~\cite{Medina:2007hz}, but the implementation of fermions in our model
benefits from the additional flexibility afforded by the general non-linear sigma model method.  In particular, we shall 
let all of the fermion fields transform as non-trivial representations of the global $SO(5)_{0}$ symmetry at site 0 only,
and as non-trivial representations of the corresponding gauge symmetries, $SU(2)_{0L}\times U(1)_{0R}$.

For each generation of quarks in the standard model, we will have three multiplets of $SO(5)_0$, ($\psi^A$, $\psi^B$, $\psi^C$), 
one each for the left-handed quark doublet $Q_L$, the right-handed up quark $u_R$, and the right-handed down quark $d_R$,
respectively.\footnote{Due to our unfortunate choice of notation, we will
be using the subscripts $L$ and $R$ to label the chirality of the fermion fields, as well as the two gauged subgroups of $SO(5)$.  When applied to a fermion field, the subscripts always denote the chirality.  Everywhere else they label the subgroup of $SO(5)$. }    The multiplets are Dirac 
multiplets, in that each comes in a right-handed and left-handed pair, 
\begin{equation}
\psi\,\equiv\,\left(
\begin{array}{c}
\psi_L\\
\psi_R
\end{array}
\right)\,,
\end{equation}
{\em except} that the standard model fields 
within the multiplet are missing
their Dirac partners.  For example, the $Q_L$ field resides in the multiplet
$\psi^A_L$, which transforms as the fundamental 5 of $SO(5)$, while the corresponding $\psi^A_R$ multiplet is
missing the $Q_R$ field.  In terms of component fields we have
\begin{equation}
\psi_L^A\,=\, \left(\begin{array}{c}
				Q\\
				\chi\\
				u
		           \end{array}\right)^A_L\,, \quad 
\psi_{R}^A\,=\, \left(\begin{array}{c}
				0 \\
				\chi\\
				u
		           \end{array}\right)^A_{R}\,,\label{eq:psia}
\end{equation}
where
\begin{equation}
Q\,=\, \left(\begin{array}{c}
				Q^{u} \\
				Q^d
		           \end{array}\right) \quad \mathrm{and} \quad 
\chi\,=\, \left(\begin{array}{c}
				\chi^{y} \\
				\chi^u\\
		           \end{array}\right)\,
\end{equation}
transform as doublets under $SU(2)_{0L}$ and $u$ transforms as a singlet.
Under $U(1)_{0R}$ the fields transform with a charge given by $Y=T^3_R+q_X$, where $q_X=+2/3$ for quarks and
$q_X=0$ for leptons\footnote{In the extra-dimensional gauge-Higgs model the charge $q_X$ arises from the extra $U(1)_X$ bulk gauge symmetry.  In our model, we are free to give the
fermion fields any charge $Y$ under the $U(1)_{0R}$, and so $q_X$ corresponds to the difference
between $Y$ and the canonical charge $T^3_R$.}.
In this way, we find that the electromagnetic charge of each component field is given by
\begin{equation}
q_{_{EM}}\ =\ T^3_L+T^3_R+q_X\ =\ T^3_L+Y\ ,
\label{eq:emcharge}
\end{equation}
a result which holds for the component fields in each $SO(5)$ multiplet.
Throughout this paper, we will use the
symbols $y$, $u$, and $d$ to indicate the electromagnetic charges of the fields by $q_{_{EM}}(y)=+5/3$, $q_{_{EM}}(u)=+2/3$,
and $q_{_{EM}}(d)=-1/3$.  

The right-handed up quark field $u_R$ resides in the multiplet $\psi^B_R$, which also transforms as the fundamental 5 of
$SO(5)$, and has a corresponding Dirac partner multiplet $\psi^B_L$, which is missing the $u_L$ field.  In terms of component fields we have
\begin{equation}
\psi_L^B\,=\, \left(\begin{array}{c}
				Q\\
				\chi\\
				0
		           \end{array}\right)^B_L\,,\quad
\psi_{R}^B\,=\, \left(\begin{array}{c}
				Q \\
				\chi\\
				u
		           \end{array}\right)^B_{R}\,.\label{eq:psib}
\end{equation}
As with the previous multiplets, the $Q$ and $\chi$ components transform as doublets under $SU(2)_{0L}$, the $u$ component
transforms as a singlet, and all component fields transform with charge $Y=T^3_R+q_X$
under $U(1)_{0R}$.

Finally, the right-handed down quark field $d_R$ resides in the multiplet $\psi^C_R$, which transforms as the adjoint 10 of
$SO(5)$, and has a corresponding Dirac partner multiplet $\psi^C_L$, which is missing the $d_L$ field.  In terms of component fields we have
\begin{eqnarray}
\psi_{L}^C&=& \frac{1}{\sqrt{2}}\left(\begin{array}{ccccc}
				-u^-& \phi^{y} & 0 & 0 & Q^{u} \\
				\phi^{d} & -u^+ & 0  & 0 & Q^{d} \\
				-y & 0 & u^+  & \phi^y & \chi^{y} \\
				0 & -y & \phi^d & u^- & \chi^{u} \\
				\chi^{u} & -\chi^y & -Q^d  & Q^u & 0 
		           \end{array}\right)^C_{L},\nonumber\\
		           \psi_{R}^C&=& \frac{1}{\sqrt{2}}\left(\begin{array}{ccccc}
				-u^-& \phi^{y} & -d & 0 & Q^{u} \\
				\phi^{d} & -u^+ & 0  & -d & Q^{d} \\
				-y & 0 & u^+  & \phi^y & \chi^{y} \\
				0 & -y & \phi^d & u^- & \chi^{u} \\
				\chi^{u} & -\chi^y & -Q^d  & Q^u & 0 
		           \end{array}\right)^C_{R},
\end{eqnarray}
where
\begin{equation}
u^\pm\,=\,\frac{1}{\sqrt{2}}\left(u\pm\phi^u\right)\,.
\end{equation}
Under $SU(2)_{0L}$,
the fields $\phi$ transform as triplets, the fields $Q$ and $\chi$ transform as doublets, and the fields $y$, $u$, and $d$
transform as singlets.  Under $U(1)_{0R}$ the fields transform with a charge given by $Y=T^3_R+q_X$ (with $T^3_R$ in
the adjoint representation for $\psi^C$), so that 
Eq.~(\ref{eq:emcharge}) holds for all fields.   

The Lagrangian density for the fermion fields with Dirac masses can be written
\begin{eqnarray}
{\cal L}_{\rm Dirac}&=& i\bar{\psi}^A\sla{D}\psi^A-\lambda_Af\bar{\psi}^A\psi^A+i\bar{\psi}^B\sla{D}\psi^B-\lambda_Bf\bar{\psi}^B\psi^B\nonumber\\
&&
+\,i\,{\rm tr}\left(\bar{\psi}^C\sla{D}\psi^C\right)-\lambda_Cf{\rm tr}\left(\bar{\psi}^C\psi^C\right) 
\ ,
\label{eq:lagrangebulk}
\end{eqnarray}
where the covariant derivatives are
\begin{eqnarray}
D^\mu\psi^{(A,B)}&=&\left[\partial^\mu
				- i g_{{0L}} W_{{0L}}^{a\mu} T_L^a 
				- i g_{{0R}} B_{{0R}}^{\mu} \left(T^3_R+q_X\right) \right]\psi^{(A,B)} 
\nonumber\\
D^\mu\psi^C&=&\partial^\mu \psi^C      
				- i g_{{0L}} W_{{0L}}^{a\mu} \left[T_L^a, \psi^C\right]
				- i g_{{0R}} B_{{0R}}^{\mu} \left(\left[T^3_R,\psi^C\right]+q_X \psi^C\right)
\,.
\label{eq:covderivferm}
\end{eqnarray}
With this Lagrangian all $\psi^A$ fields have a Dirac mass $M_A=\lambda_Af$, all 
$\psi^B$ fields have a Dirac mass $M_B=\lambda_Bf$, and all $\psi^C$ fields have a Dirac mass $M_C=\lambda_Cf$,
{\em except} for the fields with missing partners, which are massless.  For each generation of quarks 
there will be five heavy charge +5/3 fermions: one with mass $M_A$,
one with mass $M_B$ and three with mass $M_C$.  There will be three heavy charge -1/3 fermions: one with mass $M_B$
and two with mass $M_C$.  There will be eight heavy charge +2/3 fermions: two with mass $M_A$, two with mass $M_B$ 
and four with mass $M_C$.  The fields $Q^{A}_{L}$, $u^{B}_{R}$, and $d^{C}_{R}$
remain massless at this point.  

Let us consider how to give the light fermions a mass, by noting the symmetries of the Dirac mass terms in Eq.~(\ref{eq:lagrangebulk}).
They are written to appear symmetric under the $SO(5)_0$ transformation $\psi^{(A,B)}\rightarrow U_0\psi^{(A,B)}$
and $\psi^C\rightarrow U_0\psi^CU_0^\dagger$; however, this symmetry is explicitly broken
by the missing partners in the $SO(5)$ multiplets.  On the other hand, the $SO(5)_1$ symmetry
is preserved by default.  In addition, there is a global $U(1)$ symmetry associated with each of the $\psi^A$, $\psi^B$, and
$\psi^C$ fields, which must be broken to give the light fermions a mass.

We can create objects that transform under the $SO(5)_1$ symmetry, by multiplying the complete fermion multiplets by
the $\Sigma$ field:  $\psi^{A\prime}_L=\Sigma^\dagger\psi^A_L$, $\psi^{B\prime}_R=\Sigma^\dagger\psi^B_R$,
and $\psi^{C\prime}_R=\Sigma^\dagger\psi^C_R\Sigma$.  Since the $SO(5)_1$ symmetry is broken explicitly by the
gauge interactions to $O(4)_1$, we can include this breaking by projecting onto $O(4)$ invariant subspaces, using 
the $O(4)$-invariant vector,
\begin{eqnarray}
		E \  =\ 	    \left(\begin{array}{rrrrr}
						0 \\
						0 \\
						0 \\
						0 \\
						1 \\	
		                             \end{array}\right)\ 
\end{eqnarray}
It is useful to think of this vector as a spurion field which transforms as $E\rightarrow U_1E$ under the $SO(5)_1$
transformation.  In this way, we can write three Yukawa terms for the fermions that have the $SO(5)_1$ symmetry broken
purely by the vector $E$.  They are
\begin{eqnarray}
{\cal L}_{\rm Yukawa}&=& -\left[\lambda_1f\left(\bar{\psi}_{L}^A\Sigma \right) EE^\dagger\left(\Sigma^\dagger \psi_{R}^B\right)
+\sqrt{2}\lambda_2f\left(\bar{\psi}_{L}^A\Sigma\right) \left(1-EE^\dagger\right)\left(\Sigma^\dagger \psi_{R}^C\Sigma\right)E\right.\nonumber\\
&&\left.\quad\,
+\lambda_3f\left(\bar{\psi}_{L}^A\Sigma\right) \left(1-EE^\dagger\right)\left(\Sigma^\dagger \psi_{R}^B\right)
+{\rm h. c.} \right]\nonumber\\
&=& -\left[\lambda_1f\left(\bar{\psi}_{L}^A\Sigma \right) EE^\dagger\left(\Sigma^\dagger \psi_{R}^B\right)
+\sqrt{2}\lambda_2f\left(\bar{\psi}_{L}^A\psi_{R}^C\Sigma\right)E\right.\\
&&\left.\quad\,
+\lambda_3f\left(\bar{\psi}_{L}^A\Sigma\right) \left(1-EE^\dagger\right)\left(\Sigma^\dagger \psi_{R}^B\right)
+{\rm h. c.} \right]\ ,\nonumber
\label{eq:lagrangebrane}
\end{eqnarray}
where we have used the $SO(5)$ transformation properties of the adjoint representation to simplify the second term.
Note that these three terms correspond directly to the three ``brane'' mass terms in the
 5-dimensional $SO(5)\times U(1)_X$ Gauge-Higgs model of Ref.~\cite{Medina:2007hz}.  In addition we note that the Yukawa terms of Eq.~(\ref{eq:lagrangebrane}) preserve the $SO(5)_0$ symmetry, while the Dirac mass terms of Eq.~(\ref{eq:lagrangebulk}) preserve the $SO(5)_1$ symmetry, so that the fermion interactions also exhibit the collective symmetry breaking that is necessary to cancel the 
 one-loop quadratic divergences to the Higgs potential.

According to Refs.~\cite{Carena:2007ua,Carena:2006bn}, the term with $\lambda_3$ results in a large negative correction to the $T$ parameter in extra-dimensional models.  Furthermore, we can forbid this term if we assume that the terms that simultaneously break the $SO(5)_1$ 
and the global $U(1)$'s in the fermion sector must be proportional to $E$.  Thus,
we will follow the lead of Ref.~\cite{Medina:2007hz} and set $\lambda_3=0$.  Expanding in terms of component fields, we obtain
\begin{eqnarray}
{\cal L}_{\rm Yukawa}&=& -\left[\frac{isc\lambda_1f}{\sqrt{2}|H|}\Bigl(\bar{Q}_{L}^{A}H\Bigr)u_{R}^{B}
-\frac{is\lambda_2f}{\sqrt{2}|H|}\Bigl(\bar{Q}_{L}^{A}\tilde{H}\Bigr)d_{R}^{C}+\dots
+{\rm h. c.} \right]\,,
\label{eq:lagrangebraneII}
\end{eqnarray}
which contains the same Yukawa terms for the light fermions as in the standard model.
If we assume that $\lambda_{(1,2)}\ll\lambda_{(A,B,C)}$, then this results in masses for the up and down quarks of 
\begin{eqnarray}
M_u&\approx&\lambda_1v/\sqrt{2}\nonumber\\
M_d&\approx&\lambda_2v/\sqrt{2}\label{eq:fermionmass}\,,
\end{eqnarray}
while the heavy fermions get only small shifts from their masses of $M_A$, $M_B$, $M_C$.
In general, $\lambda_1$ and $\lambda_2$ will be matrices in generation space, leading to weak mixing and the CKM matrix.

The only quark for which the approximation $\lambda_{1}\ll\lambda_{(A,B,C)}$ may not hold is the top quark.  If we take
$\lambda_{1}$ for the top quark sector of the same order as $\lambda_{(A,B,C)}$ we find that the charge +2/3 fermions of $\psi_C$ and one
linear combination of each of the charge +2/3 fermions of $\psi_A$ and $\psi_B$ have mass eigenvalues unaffected by the Yukawa term.
The remaining three linear combinations mix due to the Yukawa term and have masses, to leading nonzero order in $v/f$, of
\begin{eqnarray}
M_t&\approx&\lambda_tv/\sqrt{2}\nonumber\\
M_{T_A}&\approx&\sqrt{\lambda_A^2+\lambda_{1}^{2}}f\label{eq:topmass}\\
M_{T_B}&\approx&\lambda_Bf\nonumber\,,
\end{eqnarray}
where we have defined
\begin{equation}
\frac{1}{\lambda_t^2}\,=\,\frac{1}{\lambda_{1}^{2}}+\frac{1}{\lambda_A^2}\,.\label{eq:lambdatop}
\end{equation}
We see that even for $\lambda_{1}$ not small, the top quark mass is down by a factor of $v/f$ compared to the heavy quarks.
It is possible to obtain these three mass eigenvalues exactly as the solution of a cubic characteristic equation.   The three
masses
are plotted as a function of $v/f$ in Fig.~\ref{fig:Tmass}.  
More details of the fermion masses and mixings in the top quark sector are given in Appendix~\ref{sec:fermionmass}.

\EPSFIGURE[t]{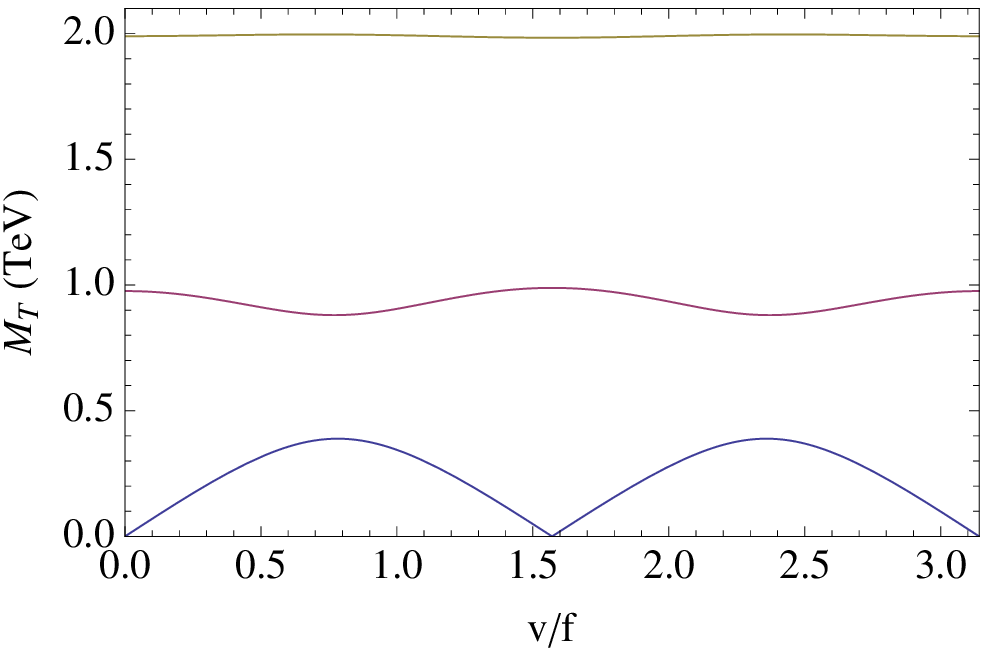,width=0.6\textwidth}
{Charged +2/3 fermion masses, in the top quark sector, as a function of $v/f$, for 
$f=1$ TeV, $\lambda_A=\lambda_1=\sqrt{2}\lambda_t$ and $\lambda_B=0.981\lambda_t$.
The curves from top to bottom are $M_{T_A}$, $M_{T_B}$, and $M_t$.
\label{fig:Tmass}}

\section{Effective Potential}
\label{sec:potential}

In our model, the vacuum expectation value of the Higgs doublet is driven entirely
by the radiatively-produced effective potential.  The potential depends on 7 independent parameters: $\{f, g_1, g_{0L}, g_{0R},\lambda_A,\lambda_B,\lambda_1\}$.  Here, we have
chosen to equate the gauge couplings at site 1: $g_1=g_{1L}=g_{1R}$.  The fermion
parameters $\lambda_A$, $\lambda_B$, and $\lambda_1$ are those for the third-generation quark sector.  We note that the additional
fermion parameters $\lambda_2$ and $\lambda_C$ can be neglected in the limit of zero
bottom quark mass; $\lambda_2$
is directly proportional to the bottom quark mass, while the heavy states in the $\psi^C$
multiplet do not mix in this limit.  Finally, we must
include a cutoff $\Lambda$ for our theory.  Using naive dimensional analysis, we choose this 
to be proportional to the symmetry-breaking scale $f$ by $\Lambda=4\pi f$.

The seven parameters listed above are not entirely unconstrained, since we must 
recover the standard model at low energies.  In particular we must recover the electroweak
gauge couplings $g\equiv g_L$ and $g^\prime\equiv g_R$, the top Yukawa coupling $\lambda_t\equiv\sqrt{2}M_t/v$, and the
Higgs vacuum expectation value $v$.  This results in four constraints on the above parameters.
Three of these relations have been given previously in Eqs.~(\ref{eq:gLgR})
and Eq.~(\ref{eq:lambdatop}).  Using Eqs.~(\ref{eq:gLgR}), it is possible to treat $g_1$
as independent, while fixing $g_{0L}$ and $g_{0R}$ by the relations
\begin{eqnarray}
		\frac{1}{g_{0L}^2}&=&\frac{1}{g_L^2}-\frac{1}{g_{1}^2}\nonumber\\
		\frac{1}{g_{0R}^2}&=&\frac{1}{g_R^{2}}-\frac{1}{g_{1}^2}\,.\label{eq:constraints}
\end{eqnarray}
Note that these equations imply that $g_1>g_{L,R}$.  We impose Eq.~(\ref{eq:lambdatop})
by defining a mixing angle in the top sector,
\begin{equation}
\sin\theta_t\ =\ \frac{\lambda_1}{\sqrt{\lambda_1^2+\lambda_A^2}}\ ,\label{eq:thetat}
\end{equation}
so that the top mass parameters are given in terms of $\theta_t$ by 
$\lambda_A=\lambda_t/\sin\theta_t$ and $\lambda_1=\lambda_t/\cos\theta_t$.  
The fourth constraint is that the minimum of the effective potential for the Higgs doublet
is at $\langle|H|\rangle=v/\sqrt{2}$.  In the following, we find it convenient to
choose the set $\{f, g_1,\sin\theta_t\}$ as our free parameters, while varying $\lambda_B$
to minimize the effective potential at the correct value of $v$.

The gauge and fermion contributions to the Higgs potential are generated at the one-loop level
and can be expressed by the formulae of Coleman and Weinberg~\cite{Coleman:1973jx}.  Because of the collective symmetry breaking, there are no quadratic divergences at this order; however, there are logarithmic divergences, which must be cutoff at the scale $\Lambda=4\pi f$.    The Coleman-Weinberg potential for our model can be written 
\begin{equation}
V\ =\ V_{\rm gauge}\,+\,V_{\rm fermion}\ ,
\end{equation}
where
\begin{eqnarray}
V_{\rm gauge}&=&\frac{3}{64 \pi^2}\left\{2~\textnormal{Tr}\left[{\cal{M}}_{CC}^4(\Sigma) \textnormal{ln}\left(\frac{{\cal{M}}_{CC}^2(\Sigma)}{\Lambda^2}\right)\right]+\textnormal{Tr}\left[{\cal{M}}_{NC}^4(\Sigma) \textnormal{ln}\left(\frac{{\cal{M}}_{NC}^2(\Sigma)}{\Lambda^2}\right)\right]\right\}\,,\nonumber\\ \label{eq:VCW}
V_{\rm fermion}&=&-\frac{3}{16 \pi^2}\textnormal{Tr}\left[\left({\cal{M}}^\dagger{\cal{M}}_{\rm top}(\Sigma)\right)^2\textnormal{ln}\left(\frac{{\cal{M}}^\dagger{\cal{M}}_{\rm top}(\Sigma)}{\Lambda^2}\right)\right]\,,
\end{eqnarray}
where ${\cal{M}}_{CC}^2$, ${\cal{M}}_{NC}^2$, and ${\cal{M}}_{\rm top}$ are given in 
the appendices in Eq.~(\ref{eq:CCmatrix}), Eq.~(\ref{eq:NCmatrix}), 
and Eq.~(\ref{eq:topmatrix}), respectively.  In general, the logarithm of the cutoff, $\ln{\Lambda^2}$,
may be accompanied by a scheme-dependent additive constant, which can only be
determined within the high-energy completed theory.  In this paper, we will set
these to zero.

We are now ready to explore the parameter space of the Coleman-Weinberg potential.  Using the masses $M_W$, $M_Z$, $M_t$ and the Fermi constant $G_F$ as inputs, we impose the constraints with $g_L^2=.426$, $g_R^2=.122$, $\lambda_t^2=.990$, and require a minimum of the potential at $v=246$ GeV.  
We consider the following range of parameters:
\begin{eqnarray}
		.5 \leq &g_1^2& \leq 4\pi \nonumber\\
	  	.1 \leq &\sin^2\theta_t& \leq .9 \label{eq:parameterspace}\\
		  1 {\rm\ TeV}\leq&f&\leq10 {\rm\ TeV}\,,\nonumber
\end{eqnarray}
which assumes that none of the dimensionless parameters in the set  $\{g_1, g_{0L}, g_{0R},\lambda_A,\lambda_1\}$ are too large.  Within this range of parameters,
we find that it is always possible to obtain two values of $\lambda_B$ for each
choice of $\{f, g_1,\sin\theta_t\}$ that give the correct vev.  In Figs.~\ref{fig:potential} and~\ref{fig:potentialzoom}
we plot the potential for a typical set of parameters $\{f=1$ TeV, $g_1^2=6$, $ \sin^2\theta_t=1/2\}$ with $\lambda_B=0.981\lambda_t$, that gives $v=246$ GeV and $M_H=130$ GeV.

\EPSFIGURE[t]{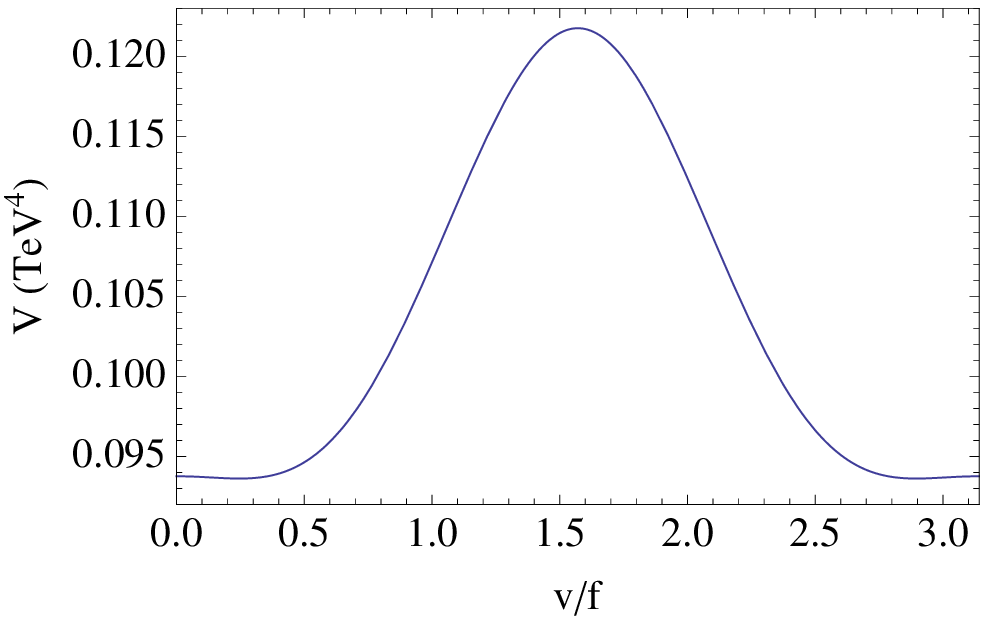,width=0.6\textwidth}
{Coleman-Weinberg Potential as a function of $v/f$, for  $g_1^2=6$,  $f=1$ TeV, $\lambda_A=\lambda_1=\sqrt{2}\lambda_t$ and $\lambda_B=0.981\lambda_t$.  This choice of parameters gives
$v=246$ GeV and $M_H=130$ GeV.
\label{fig:potential}}

\EPSFIGURE[t]{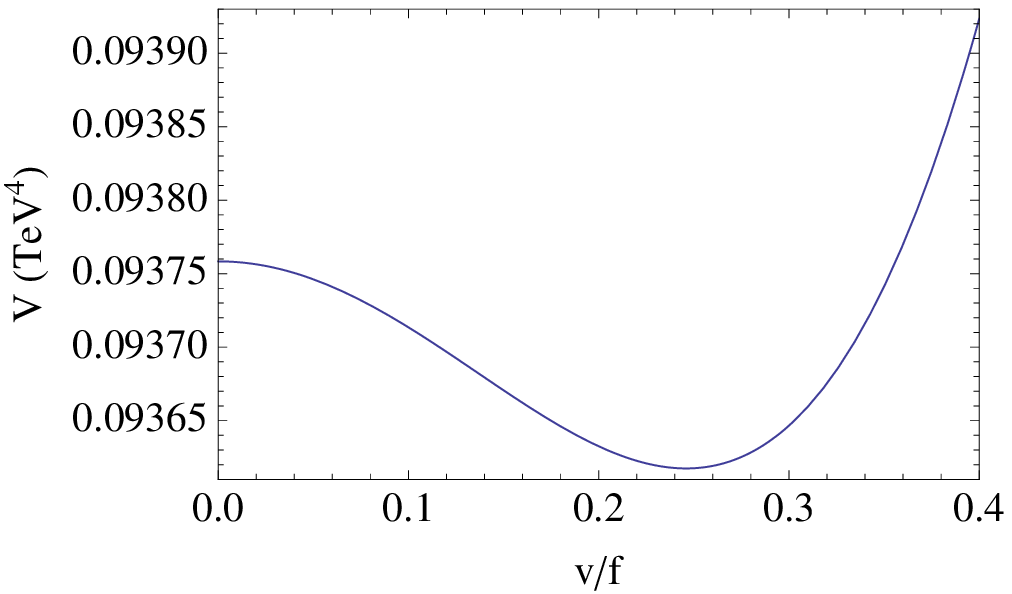,width=0.6\textwidth}
{Same as Figure~\ref{fig:potential}, but plotted  with $v/f$ ranging from $0$ to $.4$ to show the minimum in detail. 
\label{fig:potentialzoom}}

Before discussing the two different branches of solutions for $\lambda_B$ further, it is useful to consider
the Coleman-Weinberg potential, expanded for small values of the Higgs field $H$.  We have 
\begin{equation}
V\ =\ m^2H^\dagger H\,+\,\lambda(H^\dagger H)^2+\cdots\,.
\end{equation}
The full expressions for $m^2$ and $\lambda$ are given in Appendix~\ref{sec:Hpot}; however,
we find that the qualitative features of the two solutions can be understood from the dominant fermion-loop contributions to $m^2=m^2_{\rm gauge}+m^2_{\rm fermion}$.   We obtain
\begin{equation}
m^2_{\rm fermion}\ =\ \frac{3}{8\pi^2}\left\{\left(2M_{T_B}^2\lambda_1^2-M_{T_A}^2\lambda_t^2\right)
\left(\ln\frac{\Lambda^2}{M_{T_A}^2}-\frac{1}{2}\right) +\frac{2M_{T_B}^4\lambda_1^2}
{M_{T_A}^2-M_{T_B}^2}\ln\frac{M_{T_B}^2}{M_{T_A}^2}
 \right\}\ ,
\end{equation}
with $M_{T_A}^2=(\lambda_A^2+\lambda_1^2)f^2$ and $M_{T_B}^2=\lambda_B^2f^2$.

\EPSFIGURE[t]{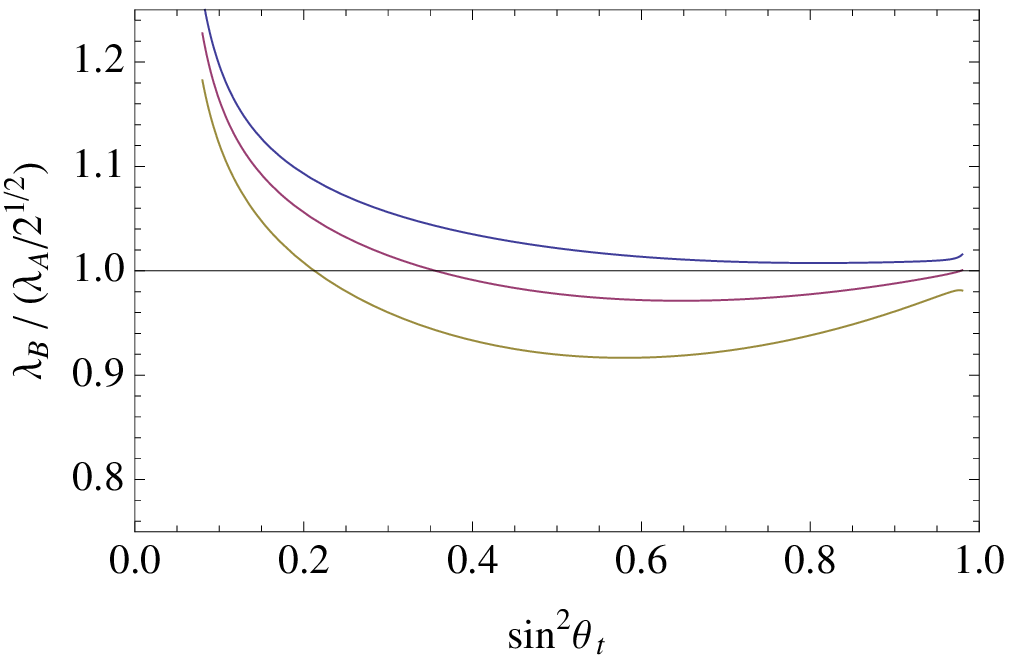,width=0.6\textwidth}
{The ``small-$M_H$'' branch of solutions for $\lambda_B/(\lambda_A/\sqrt{2})$ as a function of
$\sin^2\theta_t$ for $f=1$ TeV and for three different values of $g_1$.  From top to bottom the
three curves correspond to $g_1^2=0.5$, $g_1^2=2\pi$, and $g_1^2=4\pi$, respectively.
\label{fig:lambdaB}}

Note that $m^2_{\rm fermion}$ can be either positive or negative, due to the collaboration of the two heavy fermions.  In fact, in order to find a
Higgs vacuum expectation value with $v\ll f$, it is necessary
that the contributions to $m^2_{\rm fermion}$ cancel to some degree.  As suggested 
above, this can happen in two different ways.  Firstly, one could cancel the coefficient of 
the divergent logarithm $\ln{\Lambda^2}$, which is proportional to $(2M_{T_B}^2\lambda_1^2-M_{T_A}^2\lambda_t^2)=\lambda_1^2f^2(2\lambda_B^2-\lambda_A^2)$.  This cancels 
exactly for $\lambda_B=\lambda_A/\sqrt{2}$, giving a completely finite fermion contribution to the full Coleman-Weinberg potential at one loop.  
The choice  $\lambda_B\approx\lambda_A/\sqrt{2}$ also gives a reasonable approximation to the first (``small-$M_H$'') branch of solutions for $\lambda_B$.  This can be seen in Fig.~\ref{fig:lambdaB}, where
we plot $\lambda_B/(\lambda_A/\sqrt{2})$ for this branch as a function of $\sin^2\theta_t$ for $f=1$ TeV and for three different values of $g_1^2$.  Over most of the range of $\sin^2\theta_t$, we find $\lambda_B\approx\lambda_A/\sqrt{2}$ within 10\%.
As we shall see later in this section, the simple relation between $\lambda_A$ and $\lambda_B$ is in general modified by ultraviolet effects, but it is still possible to find a choice of $\lambda_B$ that gives $v=246$ GeV and a light Higgs boson for most of the parameter space.
The predictions for the Higgs boson mass that correspond to the solutions given here are shown in Fig.~\ref{fig:HiggsMass} for $f=1$ TeV
and $f=10$ TeV for the same three values of $g_1^2$.  For the range of parameters given in
Eq.~(\ref{eq:parameterspace}) we
find $120$ GeV$\lesssim M_H \lesssim 320$ GeV, with the lighter values of $M_H$ corresponding
to smaller values of $\lambda_A$ and larger values of $\lambda_1$.  
In particular, for $f=1$ TeV, we obtain $M_H\lesssim150$ GeV over a large range of $\sin^2\theta_t$.
Interestingly, the predictions for $M_H$ show very little
dependence on the gauge coupling $g_1$, with $M_H$ varying by only a few GeV
for $0.5\le g_1^2\le4\pi$.  Furthermore, the predictions
show only modest dependence on $f$, with $M_H$ increasing by about 40 GeV as
$f$ is increased from 1 TeV to 10 TeV.

\EPSFIGURE[t]{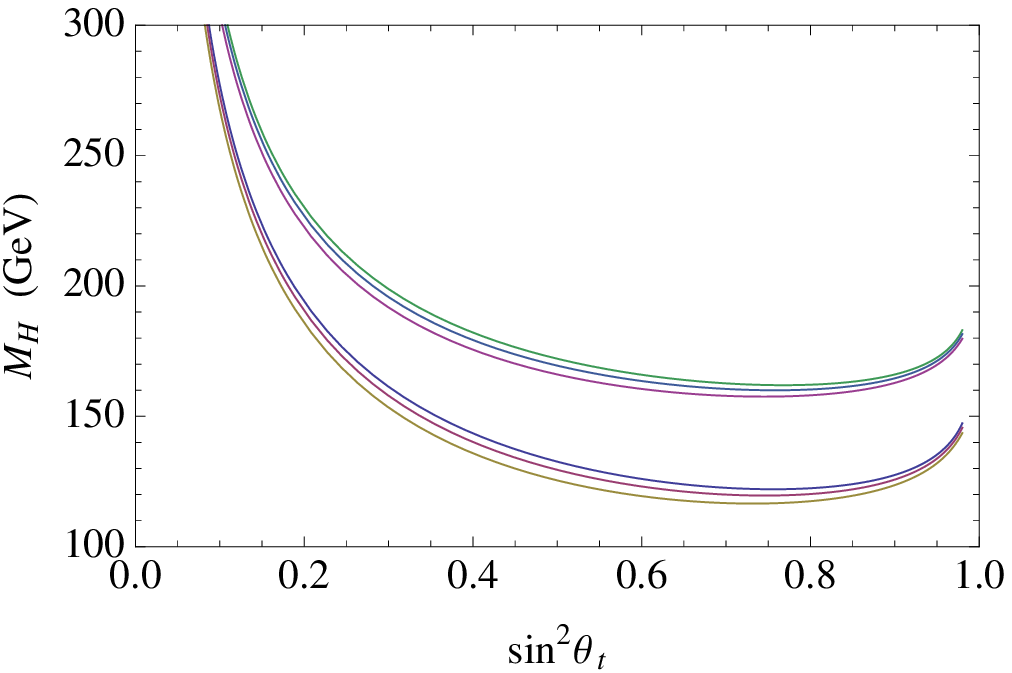,width=0.6\textwidth}
{The ``small-$M_H$'' branch predictions for the Higgs boson mass as a function of
$\sin^2\theta_t$.  The upper three curves are for $f=10$ TeV, while the lower three curves are for $f=1$ TeV.   Within each set of three, the curves correspond from top to bottom to $g_1^2=0.5$, $g_1^2=2\pi$, and $g_1^2=4\pi$, respectively.
\label{fig:HiggsMass}}

The second (``large-$M_H$'') branch of solutions for $\lambda_B$ can also be identified with a cancellation in $m^2_{\rm fermion}$.  In this case
the cancellation occurs for large $M_{T_B}$, with the result
$M_{T_B}^2\approx\Lambda^2e^{-1/2}$.  The exact solutions have  $7\lesssim\lambda_B/\lambda_t\lesssim9$, with corresponding
values of the Higgs boson mass of $380$ GeV$\lesssim M_H \lesssim 910$ GeV.
As with the other branch of solutions, we find that the values of $\lambda_B$ and $M_H$
depend mostly on $\sin^2\theta_t$, with little dependence on $g_1$ and $f$.
On the other hand, this branch of solutions is probably not satisfactory, since it
requires the mass $M_{T_B}$ of one of the heavy fermions to be of the same size as the
cutoff $\Lambda$.  In addition, this solution will be strongly affected by the inclusion of
a scheme-dependent constant, $\ln{\Lambda^2}\rightarrow\ln{\Lambda^2}+\delta_F$,
which again shows that the theory with this choice of $\lambda_B$  will be strongly influenced 
by unknown dynamics at the cutoff.  Finally, the larger values of $M_H$ obtained for this branch of solutions also makes it
less viable phenomenologically, as we will see in the next section.  For these reasons, we focus on the ``small-$M_H$'' branch
of solutions in the remainder of this paper.

One may wonder whether the ``small-$M_H$'' branch of solutions is also strongly affected by ultraviolet physics at the cutoff scale.  
For instance, if there is a different cutoff associated with the $\psi^A$ fermions and the $\psi^B$ fermions, one might expect that
the factor 
\[
\left(2M_{T_B}^2\lambda_1^2-M_{T_A}^2\lambda_t^2\right)\ln\frac{\Lambda^2}{M_{T_A}^2}\ =\ \lambda_1^2f^2(2\lambda_B^2-\lambda_A^2)\ln\frac{\Lambda^2}{M_{T_A}^2}\ ,
\]
 which is strongly canceled in this branch
of solutions, would be replaced by
\[
\lambda_1^2f^2\left(2\lambda_B^2\ln\frac{\Lambda_B^2}{M_{T_A}^2}-\lambda_A^2\ln\frac{\Lambda_A^2}{M_{T_A}^2}\right)\ .
\]
In Appendix \ref{sec:fermionmod} we present a modification of the fermion sector that leaves the fermion contribution to the one-loop Coleman-Weinberg potential for the Higgs boson finite, and has exactly the effect just described above. In this case there is an additional term in the potential,
\begin{eqnarray}
\Delta V_{\rm fermion}&=&-\frac{3}{16 \pi^2}f^4\lambda_1^2s^2\left\{2\lambda_B^2
\ln\frac{\Lambda^2}{\Lambda_{B}^{ 2}} - \lambda_A^2
\ln\frac{\Lambda^2}{\Lambda_{A}^{2}}
 \right\}\ ,\label{eq:Vmodi}
 \end{eqnarray}
which exactly cancels the dependence on the UV cutoff $\Lambda$ in $V_{\rm fermion}$ of Eq.~(\ref{eq:VCW}), exchanging it for the
dependence on the two large mass parameters, $\Lambda_A$ and $\Lambda_B$.  

\EPSFIGURE[t]{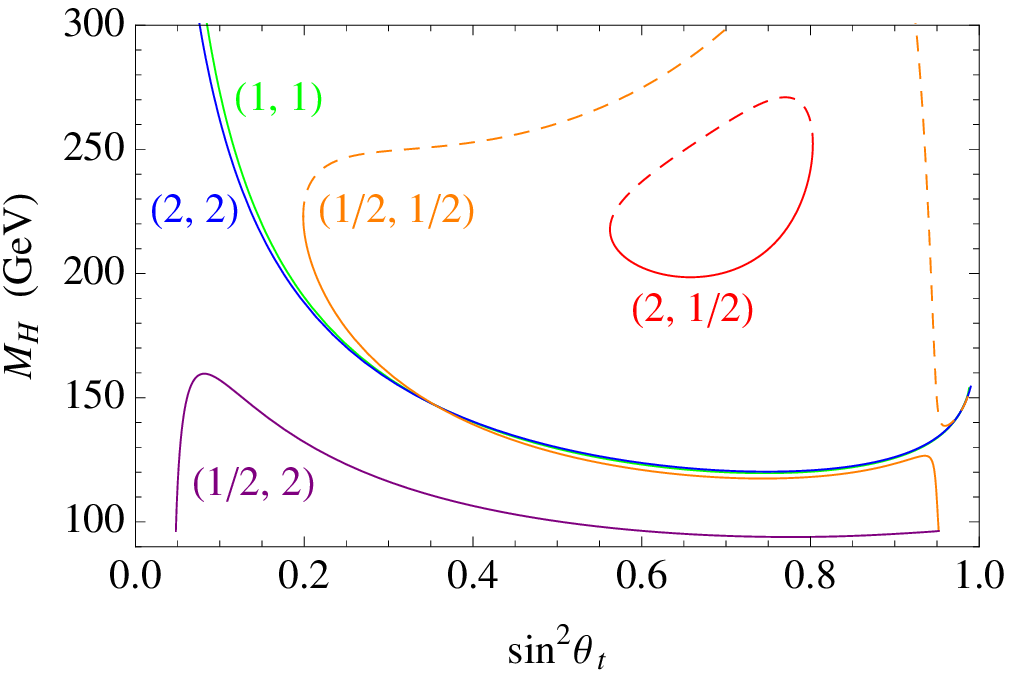,width=0.6\textwidth}
{Sensitivity of the ``small-$M_H$'' branch predictions for the Higgs boson mass to non-identical fermion cutoffs.
 All four curves are for $f=1$ TeV, $g_1^2=2\pi$.  The curves are labeled by $(\Lambda_A/\Lambda,\Lambda_B/\Lambda)$, where $\Lambda=4\pi f$.  The dashed curves are the corresponding``large-$M_H$" branch predictions for the Higgs boson mass, which lie in the
 mass-range of this plot for $\Lambda_B/\Lambda=1/2$.
\label{fig:HiggsMassSensitivity}}

For $\Lambda_A\ne\Lambda_B$, the ``small-$M_H$'' solutions now occur for
\begin{equation}
\lambda_B^2\approx\frac{\lambda_A^2}{2}\left(\frac{\ln(\Lambda_A^2/M_{T_A}^2)}{\ln(\Lambda_B^2/M_{T_A}^2)}\right)\ .\label{eq:Cutoffs}
\end{equation}
This implies that $M_{T_B}=\lambda_Bf$ is no longer completely determined by $M_{T_A}$ (or equivalently, by $\lambda_A$ or $\sin\theta_t$), since the relationship is modified by the ratio of logarithms of the unknown cutoffs, $\Lambda_A$ and $\Lambda_B$.   However, the Higgs boson mass is still strongly correlated
with the two heavy fermion masses $M_{T_A}$ and $M_{T_B}$.
In Fig.~\ref{fig:HiggsMassSensitivity} we investigate the sensitivity of the Higgs boson mass to UV effects by plotting $M_H$ as a function
of $\sin^2\theta_t$, while varying $\Lambda_A$ and $\Lambda_B$ together and independently between $\Lambda/2$ and $2\Lambda$, where
$\Lambda=4\pi f$.  We use $f=1$ TeV and $g^2=2\pi$ as representative values in this plot.  As expected, and in contrast to the ``large-$M_H$'' branch of solutions, the prediction for the Higgs mass is very insensitive to varying the scales together from $(\Lambda_A/\Lambda,\Lambda_B/\Lambda)=(1/2,1/2)$ to $(2,2)$, at least for $0.3\lesssim\sin^2\theta_t\lesssim0.9.$
On the other hand, for  $(\Lambda_A/\Lambda,\Lambda_B/\Lambda)=(1/2,2)$ the predictions for $M_H$ decrease by about 25-40 GeV,
while for $(\Lambda_A/\Lambda,\Lambda_B/\Lambda)=(2,1/2)$ the predictions for $M_H$ increase by about 80 GeV.  For this latter choice
of cutoffs, it can be seen from the figure that a solution for $v=246$ GeV is only obtained for $0.6\lesssim\sin^2\theta_t\lesssim0.8.$
This is related to the fact that the ``large-$M_H$'' solutions decrease in energy for smaller $\Lambda_B$, as displayed by the
dashed curves in Fig.~\ref{fig:HiggsMassSensitivity}.  The sensitivity of the Higgs boson mass to non-identical fermion cutoffs can be understood
largely in terms of the residual dependence of the Higgs quartic coupling $\lambda$ on the heavy fermion mass ratio $M_{T_B}/M_{T_A}$ (see Eq.~(\ref{eq:lambdafermion}) in Appendix~\ref{sec:Hpot}), which in turn is affected by Eq.~(\ref{eq:Cutoffs}).  Thus, fixing the two heavy fermion masses largely determines the Higgs boson mass, with larger values of $M_H$  correlated with larger values of $M_{T_B}/M_{T_A}$ for a given $\sin^2\theta_t$.
In addition, we note that over much of the parameter space the predicted Higgs boson mass is still below 200 GeV for a significant portion of the range of $\sin^2\theta_t$.

To conclude this section, we comment on the size of the fine-tuning\footnote{We have not considered here the ``hidden'' fine-tuning necessary to maintain the global symmetry of the fermion couplings against non-symmetric running, as discussed in Ref.~\cite{Grinstein:2009ex}.  Our model, like other Little Higgs models, is not obviously immune to this effect.} that is needed in this model to obtain a Higgs vacuum expectation value with $v^2\ll f^2$.
We have investigated this issue by analyzing the fine-tuning of $v^2$ with respect to the parameters $p_i\in\{g_{1L}, g_{1R}, g_{0L}, g_{0R},\lambda_A,\lambda_B,\lambda_1,\Lambda_A,\Lambda_B\}$, where the fine-tuning with respect to $p_i$ is defined by $\Delta_{p_i}=(p_i/v^2)(\partial v^2/\partial p_i)$, following Barbieri and Giudice~\cite{Barbieri:1987fn}.   We then let the total fine-tuning be the combination of each of the separate fine-tunings in quadrature, $\Delta=(\sum_i \Delta_{p_i}^2)^{1/2}$, subject to the constraints, (\ref{eq:lambdatop}) and (\ref{eq:constraints}).  Details of the formalism that we have followed can be found in Ref.~\cite{Casas:2005ev}.  For $f=1$ TeV, $g_{1L}^2=g_{1R}^2=2\pi$, and $\Lambda_A=\Lambda_B=\Lambda=4\pi f$, we find values of $\Delta$ of $\sim100-140$ for Higgs masses
between 120 and 160 GeV, with the dominant contributions coming from $\Delta_{\lambda_B}$ and $\Delta_{\lambda_A}$ (including the associated constraint).
These values are comparable to the minimium values obtained for the Simplest~\cite{simplest} and Littlest~\cite{Arkani-Hamed:2002qy} Little Higgs models, which are displayed in Fig.~13 of Ref.~\cite{Casas:2005ev}. The fact that the fine-tuning is of similar size in our model is not surprising, since all of the Little Higgs models
considered in Ref.~\cite{Casas:2005ev}, as well as our model, contain the exact same large negative contribution to $m^2$ from a heavy
partner of the top quark:
\begin{equation}
\delta m^2\ =\ -\frac{3\lambda_t^2}{8\pi^2}M_{T}^2
\ln\frac{\Lambda^2}{M_{T}^2}\ .
\end{equation}
The different models have different mechanisms for (partially) canceling this term to obtain a light Higgs boson, but since the size of this term is comparable in all of the Little Higgs models considered, one would expect the amount of fine-tuning to also be comparable.  We do note, however, that the amount of fine-tuning can be reduced in our model if we allow $\Lambda_A$ and $\Lambda_B$ to become as low as $\Lambda/3$, which reduces
the logarithmic enhancement of the above term.   In this case we can obtain
values of $\Delta$ of $\sim40-50$ for Higgs masses between 120 and 160 GeV, with the dominant contributions now coming from $\Delta_{\Lambda_A}$ and $\Delta_{\Lambda_B}$.  These amounts of fine-tuning are typically below the values for the Minimal Supersymmetric Standard Model in the same range of Higgs masses as shown in Fig.~13 of Ref.~\cite{Casas:2005ev}.  Given the ambiguities in precisely quantifying the amount of fine-tuning, we prefer to be conservative in our conclusions from this investigation, taking away from it
simply that the amount of fine-tuning in our model is comparable and typically no worse than other Little Higgs models.

\section{Electroweak Constraints}
\label{sec:constraints}

The first place to consider for testing the experimental viability of any beyond-the-standard-model theory is in constraints from electroweak precision measurements.  In our model, 
the electroweak observables receive tree-level corrections from the new gauge fields. 
In fact, although the standard model light fermions couple to all of the massive gauge fields, which are
mixtures of the gauge fields at site 0 and site 1, they are only charged under the $SU(2)_{0L}\times U(1)_{0R}$ gauge symmetry.  As a result, the corrections to low-energy observables occur only through electroweak gauge current correlators, and are thus ``universal'' in the sense of Barbieri {\em et al.}~\cite{Barbieri:2004qk}. The correlators can be easily computed from the quadratic Lagrangian by inverting the subset of the propagator matrix involving the site-0 fields only. This leads to the following expressions for the electroweak parameters~\cite{Barbieri:2004qk}, to leading order in $v^2/ f^2$:
\begin{eqnarray}
\hat{S} &=& \frac{v^2}{4f^2}\left(\sin^2\phi_L\ +\ \cot^2\theta\ \sin^2\phi_R\right)\label{eq:S} \\
\hat{T} &=& 0 \label{eq:T} \\
Y &=& \frac{v^2}{2f^2}\cot^2\theta\ \sin^4\phi_R \label{eq:Y} \\
W &=& \frac{v^2}{2f^2}\sin^4\phi_L\ . \label{eq:W}
\end{eqnarray}
Here $\sin\phi_L=g_L/g_{1L}$ and $\sin\phi_R=g_R/g_{1R}$ are defined in Eq.~(\ref{eq:phiL}) and Eq.~(\ref{eq:phiR}), respectively, and $\theta$ is the weak mixing angle defined in Eq.~(\ref{eq:weakmixing}).
We can express the couplings $g_{L}\equiv g$ and $g_{R}\equiv g^\prime$ in terms of
$\alpha(M_Z^2)$, $M_Z$, and $G_F$, and in addition
we have $v^2=1/(\sqrt{2}G_F)$ and
\begin{eqnarray}
\sin 2\theta =\left[\frac{4\pi\alpha(M_Z^2)}{\sqrt{2}G_F M_Z^2}\right]^{1/2} \ .
\end{eqnarray}
Notice that the corrections to the electroweak observables are not oblique, since nonzero values for $Y$ and $W$ signal the presence of direct corrections, corresponding to four-fermion operator exchanges at zero momentum~\cite{Barbieri:2004qk,Chivukula:2004af}. Notice also that the custodial symmetry of the model ensures that $\hat{T}=0$ at tree-level.

The observables of Eqs.(\ref{eq:S})-(\ref{eq:W}) depend on three unknown parameters: $f$, $g_{1L}$ and $g_{1R}$. In an $O(4)_1$ theory the two couplings are identical, $g_{1L}=g_{1R}\equiv g_1$, and thus we can nicely constrain the model in a two-parameter space $(f,g_1)$. 
The global fit in Ref.~\cite{Barbieri:2004qk} to the experimental data implies that a heavy Higgs boson is only compatible with positive $\hat{T}$; therefore,
we only consider the ``small-$M_H$'' branch of solutions.
The combined experimental constraints on $\hat{S}$, $Y$, and $W$, taken from Ref.~\cite{Barbieri:2004qk} with the light Higgs fit, give the bounds of Fig.~\ref{fig:bounds}, where the colored area is excluded at the 95\% confidence level. The representative values used in the plots in the previous sections, $f=1$ TeV and $g_1^2=6$, are within the allowed region. The bounds in Fig.~\ref{fig:bounds} are not expected to be strongly affected by loop corrections; however, there may be constraints on the heavy top quark sector coming from one loop contributions to the $\hat{T}$ parameter.  An
analysis of these contributions is currently underway~\cite{flsy}.
\EPSFIGURE[t]{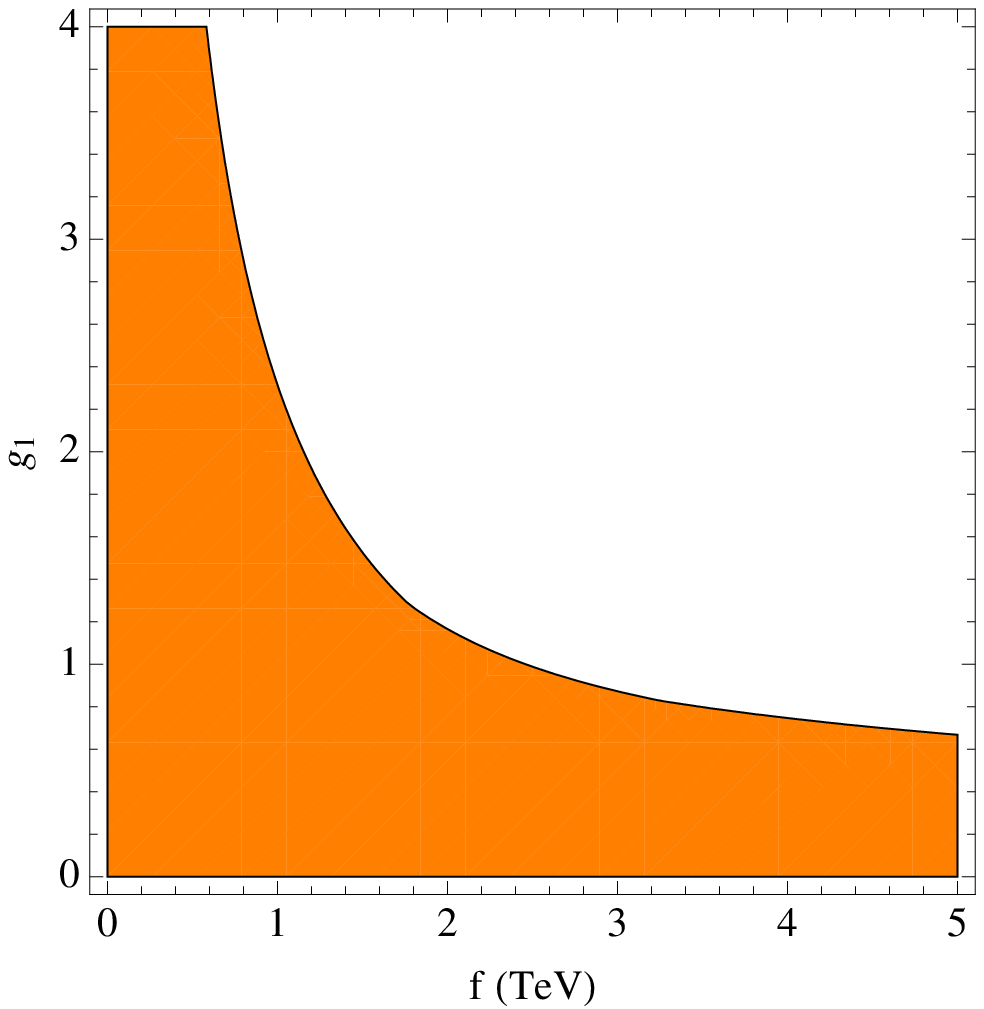,width=0.6\textwidth}
{Bounds on $g_1$ and $f$ from combined experimental constraints on $\hat{S}$, $Y$, and $W$, at the 95\% confidence level.
\label{fig:bounds}}

Finally, we must comment on the fact that the couplings of the standard model fermions to the
gauge boson eigenstates, given in Eqs.~(\ref{eq:lagrangebulk}) and (\ref{eq:covderivferm}), are not unique, in the sense that one can always add operators that correspond to renormalizations of
the broken currents:
\begin{eqnarray}
{\Delta\cal L}_{\rm Dirac}&=& i\kappa_A\bar{\psi}_L^A\left(\Sigma\sla{D}\Sigma^\dagger\right)\psi_L^A+ i\kappa_B\bar{\psi}_R^B\left(\Sigma\sla{D}\Sigma^\dagger\right)\psi_R^B\nonumber\\
&&
+\,i\kappa_{C_1}\,{\rm tr}\left[\bar{\psi}_R^C\left(\Sigma\sla{D}\Sigma^\dagger\right)\psi_R^C\right]+\,i\kappa_{C_2}\,{\rm tr}\left[\bar{\psi}_R^C\gamma^\mu\psi_R^C\left(D_\mu\Sigma\right)\Sigma^\dagger\right]\ .
\label{eq:kappas}
\end{eqnarray}
In the main discussion we have assumed that all of the fermions act as fundamental point
particles, charged only under the $SU(2)_{0L}\times U(1)_{0R}$ gauge symmetry.  In that case, the $\kappa_i$ coefficients would arise only perturbatively through loop diagrams, and we
can assume them to be small.  On the other hand, it is possible to imagine a more general scenario where these coefficients are of order one.  In fact, in the deconstruction of the
gauge-Higgs model of Ref.~\cite{Medina:2007hz} the fundamental fields that naturally appear are actually
$\psi^{A\prime}_L=\Sigma^\dagger\psi^A_L$, $\psi^{B\prime}_R=\Sigma^\dagger\psi^B_R$,
and $\psi^{C\prime}_R=\Sigma^\dagger\psi^C_R\Sigma$, which are charged under
the $SU(2)_{1L}\times SU(1)_{1R}$ gauge symmetry.  This corresponds to the case
where $\kappa_A=\kappa_B=\kappa_{C_1}=\kappa_{C_2}=1$.  In this
case the electroweak corrections are not ``universal'', and in addition, there will be a nonzero 
contribution to $\hat{T}$.  For these reasons, we have chosen the simpler fermion implementation
of Sec.~\ref{sec:fermion}, and we assume that the $\kappa_i$ are negligible.

\section{Conclusions}
\label{sec:conclusions}

In this article, we have presented a new Little Higgs model, motivated by the deconstruction of a five-dimensional gauge-Higgs model~\cite{Medina:2007hz}.  It is based on the approximate global symmetry breaking pattern $SO(5)_0\times SO(5)_1\stackrel{f}{\rightarrow}SO(5)$,
with gauged subgroups spontaneously breaking under the pattern $[SU(2)_{0L}\times U(1)_{0R}]\times O(4)_1\stackrel{f}{\rightarrow}SU(2)_L\times U(1)_Y\stackrel{v}{\rightarrow}U(1)_{EM}$, where we have made the simplifying assumption of $g_{1L}=g_{1R}$.
The novel features of this model are these: the only physical scalar in the effective theory is the Higgs boson; 
the model contains a custodial symmetry, which ensures that $\hat{T}=0$ at tree-level; and the potential for the Higgs boson is generated
entirely through one-loop radiative corrections.  Due to the collective symmetry breaking in the model these corrections have no quadratic
divergences, depending only logarithmically on the cutoff of the effective theory.

The fact that the electroweak symmetry breaking is fully radiatively-generated is a unique and intriguing feature of this model.  In particular, it implies
that the model is  more constrained, and arguably more predictive, than other Little Higgs models.  For instance, 
if we use a single cutoff $\Lambda$ for the fermion logarithmic divergences, then once the scale $f$ is chosen
and the correct value of the Higgs boson vev, $v$, is imposed, we find that the Higgs boson mass, as well as the masses of the heavy partners of the top quark, depend almost exclusively on a single
fermion mixing parameter, $\sin^2\theta_t$. 
For the ``small-$M_H$'' branch, we find for $f=1$ TeV that the Higgs boson mass satisfies 120 GeV $\lesssim M_H\lesssim150$ GeV over most of the range of $\sin^2\theta_t$.  For $f$ raised to 10 TeV, these values increase by about 40 GeV.
If we take into account possible UV effects in the fermion sector by introducing two distinct fermion cutoffs $\Lambda_A$ and $\Lambda_B$,
we still find that the Higgs boson mass is correlated with the masses of the heavy top quark partners, and it lies below 200 GeV for much of the parameter space.

The radiative symmetry breaking is achieved in this model with an amount of fine-tuning that is of similar size as in other Little Higgs models.  The relation $v\ll f$ is obtained by a cancellation between the contributions of two different heavy top quark partners to the Higgs boson mass-squared.
Once this cancellation is achieved, the Higgs boson is automatically light in the ``small-$M_H$'' branch of solutions, with the phenomenologically-viable range of masses given above.  This contrasts with other little Higgs models, where an additional operator is included to give a large Higgs quartic 
coupling and $v\ll f$, but  a similar cancellation of contributions to $m^2$ is still necessary to keep the Higgs boson light~\cite{Casas:2005ev}.

We have analyzed the tree-level constraints on the model from electroweak precision experiments and found that the model is viable
for a reasonably large and phenomenologically-interesting range of $f$ and $g_1\equiv g_{1L}=g_{1R}$.  The model introduces a number of new states, which may be probed at the LHC.  In addition to the Higgs boson, there are two heavy neutral vector bosons and two heavy charged vector bosons, whose masses and couplings depend directly on $f$ and $g_1$.  In the third-generation fermion sector, there are eight new
heavy up-like quarks, three new heavy down-like quarks, and five new heavy charge 5/3 quarks.  The masses and mixings of some portion of these heavy top quarks
will satisfy relations required by the radiative symmetry breaking and which depend on the Higgs boson mass.  If the other generations of quarks follow the
same multiplet structure, which is probably necessary to avoid flavor-changing neutral currents, this heavy fermion zoo will be multiplied by the number of generations.  In addition, similar heavy partners for the leptons should exist.  Since the decay rates of these heavy fermions to
the SM fermions are proportional to mixing angles, which in turn are proportional to the light fermion masses, it is possible that some
of these heavy particles may have long lifetimes, with interesting decay signatures.  We expect there to be a rich phenomenology
at the LHC, which demands a more detailed study~\cite{flsy}.

\vskip .2 cm

\section*{Acknowledgments}

This work was supported by the US National
Science Foundation under grant PHY-0555544.
J.H.Y. would also like to
acknowledge the support of the U.S. National Science Foundation
under grant PHY-0555545 and PHY-0855561.

\appendix

\section{$SO(5)$ Generator Matrices}
\label{sec:matrices}

Here we give a basis for the ten $SO(5)$ generator matrices that is particularly useful for our purposes.  The $5\times5$ 
generator matrices in the standard basis are

\begin{equation}
\left(T^{ab}\right)_{ij}\ =\ \frac{-i}{\sqrt{2}}(\delta^a_{i}\delta^b_{j}-\delta^a_{j}\delta^b_{i})\ ,
\end{equation}
where $a,b=1\dots5$ (with $a<b$) are the generator labels, $i,j=1\dots5$ are the row and column indices, 
and we have chosen the normalization, ${\rm tr}\left(T^{ab}T^{cd}\right)=\delta^{ac}\delta^{bd}$,
so that the gauged $SU(2)$ sub-matrices have the conventional normalization.  

It is possible to perform a similarity transformation on these matrices, $T^\prime=S^\dagger TS$, such that two of them
are simultaneously diagonal.  For example, it is possible to diagonalize $T^{\prime12}$ and $T^{\prime34}$ by the
matrix 
\begin{eqnarray}
		S \  =\ \frac{1}{\sqrt{2}}
					    \left(\begin{array}{ccccc}
						1 & 0 & 0 & -1 & 0 \\
						i & 0 & 0 & i & 0 \\
						0 & 1 & 1 & 0 & 0 \\
						0 & i & -i & 0 & 0 \\
						0 & 0 & 0 & 0 & \sqrt{2} \\	
		                             \end{array}\right)\ .
\end{eqnarray}
Applying this similarity transformation to all of the matrices and choosing convenient linear combinations of them, we obtain
the following set of basis matrices:  $\{T^{a}_L, T^{a}_R, T^1,T^2,T^3,T^4\}$, where 
\begin{eqnarray}
		T^a_L \  =\  \left(\begin{array}{cc}
						\Biggl( I\otimes\left(\half\sigma^a\right)\Biggr) &
						\begin{array}{c}
						0\\
						0\\
						0\\
						0\\
						\end{array} \\
						\begin{array}{cccc}
						\, 0\,  &\, 0\, &\, 0\, &\, 0\, \\
						\end{array}& 0 \\	
		                             \end{array}\right)\ , \qquad	                             
		T^a_R \  =\  \left(\begin{array}{cc}
						\Biggl(-\left(\half\sigma^a\right)^T\otimes I\Biggr) &
						\begin{array}{c}
						0\\
						0\\
						0\\
						0\\
						\end{array} \\
						\begin{array}{cccc}
						\ 0\  &\  0\  &\  0\  &\  0\ \\
						\end{array}& 0 \\	
		                             \end{array}\right)\ ,      \nonumber
\end{eqnarray}
\begin{eqnarray}
		T^1 \  =\ \frac{1}{2}
					    \left(\begin{array}{ccccc}
						0 & 0 & 0 & 0 & 1 \\
						0 & 0 & 0 & 0 & 0 \\
						0 & 0 & 0 & 0 & 0 \\
						0 & 0 & 0 & 0 & 1 \\
						1 & 0 & 0 & 1 & 0 \\	
		                             \end{array}\right)\ , \qquad
		T^2 \  =\ \frac{1}{2}
					    \left(\begin{array}{ccccc}
						0 & 0 & 0 & 0 & i \\
						0 & 0 & 0 & 0 & 0 \\
						0 & 0 & 0 & 0 & 0 \\
						0 & 0 & 0 & 0 & -i \\
						-i & 0 & 0 & i & 0 \\	
		                             \end{array}\right)\ ,
		                             \label{genmatrices}
\end{eqnarray}
\begin{eqnarray}		                             
		T^3 \  =\ \frac{1}{2}
					    \left(\begin{array}{ccccc}
						0 & 0 & 0 & 0 & 0 \\
						0 & 0 & 0 & 0 & 1 \\
						0 & 0 & 0 & 0 & -1 \\
						0 & 0 & 0 & 0 & 0 \\
						0 & 1 & -1 & 0 & 0 \\	
		                             \end{array}\right)\ , \qquad
		T^4 \  =\ \frac{1}{2}
					    \left(\begin{array}{ccccc}
						0 & 0 & 0 & 0 & 0 \\
						0 & 0 & 0 & 0 & i \\
						0 & 0 & 0 & 0 & i \\
						0 & 0 & 0 & 0 & 0 \\
						0 & -i & -i & 0 & 0 \\	
		                             \end{array}\right)\ .	\nonumber	                             
\end{eqnarray}
In the above expressions, $I$ is the $2\times2$ identity matrix and $\sigma^a$ are the $2\times2$ Pauli matrices for $a=1,2,3$.

\section{Gauge Boson Masses and Mixing}
\label{sec:gaugemass}

From Eq.~(\ref{eq:gaugemass}), we can obtain the mass terms for the neutral and charged gauge bosons of the following form:
\begin{equation}
{\cal{L}_{\rm mass}} = {\cal W}^{\mu\dagger} {\cal M}_{CC}^2 {\cal W}_\mu + \frac{1}{2} {\cal Z}^{\mu\dagger} {\cal M}_{NC}^2 {\cal Z}_\mu
\,,
\label{eq:Lmass}
\end{equation}
where the vectors ${\cal W}^\mu$ and ${\cal Z}^\mu$ are given by:
\begin{eqnarray}
{\cal W}^\mu = \left(
\begin{array}{c}
W^{+\mu}_{0L} \\ 
W^{+\mu}_{1L}  \\
W^{+\mu}_{1R} 
\end{array}
\right) \,,    \quad \quad {\cal Z}^\mu = \left(
\begin{array}{c}
W^{3\mu}_{0L} \\
W^{3\mu}_{1L}  \\
W^{3\mu}_{1R}  \\
B^{\mu}_{0R} \\
\end{array}
\right) \,,
\end{eqnarray}
with $W^{\pm\mu}=(W^{1\mu}\mp i W^{2\mu})/\sqrt{2}$ for each of the $SU(2)$ groups.

\subsection{The Charged Sector}
\label{subsec:charged-mass-estates}

We first consider the charged gauge boson sector.  The mass matrix in this sector 
takes the form: 
\begin{eqnarray}
{\cal{M}}_{CC}^2 & = & 
\frac{f^2}{2}
\left(
\begin{array}{ccc} 
g_{{0L}}^2          		&\ \ -(1-a)g_{{0L}} g_{{1L}} 	&\ \ -a g_{{0L}} g_{{1R}} \\
-(1-a)g_{{0L}} g_{{1L}} 	&  g_{{1L}}^2  		& 0			\\
-a g_{{0L}} g_{{1R}} 		& 0 				& g_{{1R}}^2 
\end{array}
\right) \,.\label{eq:CCmatrix}
\end{eqnarray}
For $a=0$ this mass matrix can be diagonalized in terms of the mixing angle $\phi_L$, given by
\begin{eqnarray}
		 \sin\phi_L & = & \frac{g_{{0L}}}{\sqrt{g_{{0L}}^2+g_{{1L}}^2}} \,,\nonumber\\
		 \cos\phi_L & = & \frac{g_{{1L}}}{\sqrt{g_{{0L}}^2+g_{{1L}}^2}} \,.\label{eq:phiL}
\end{eqnarray}
Recalling the coupling $g_L$, defined in Eq.~(\ref{eq:gLgR}), this implies
\begin{eqnarray}
		 g_L & = & g_{{0L}}\cos\phi_L = g_{{1L}}\sin\phi_L \,.
\end{eqnarray}

For nonzero vacuum expectation value we can solve perturbatively in the small parameter,
\begin{eqnarray}
		a 
		&=& \sin^2\left(\frac{|H|}{\sqrt{2}f}\right)\, =\,
			\frac{|H|^2}{2f^2} - \frac{|H|^4}{12f^4} + \cdots \,,
\end{eqnarray}
There will be one light eigenstate, $W^{\pm\mu}$, which we will identify as the standard model $W^\pm$, and two
heavy eigenstates, $W^{\pm\mu}_L$ and $W^{\pm\mu}_R$.  To ${\cal O}(a^2)$, the masses are
\begin{eqnarray}
M_W^2&\approx&\frac{f^2}{2}\left[2ag_L^2-a^2g_L^2\left(\cos^2{2\phi_L}+1\right)\right]\nonumber\\
M_{W_L}^2&\approx&\frac{f^2}{2}\left[(g_{0L}^2+g_{1L}^2)-2ag_L^2+a^2\left(g_L^2\cos^2{2\phi_L}
+\frac{g_{0L}^2g_{1R}^2\sin^2{\phi_L}}{g_{0L}^2+g_{1L}^2-g_{1R}^2}\right)\right]\nonumber\\
M_{W_R}^2&\approx&\frac{f^2}{2}\left[g_{1R}^2+a^2\left(g_L^2
-\frac{g_{0L}^2g_{1R}^2\sin^2{\phi_L}}{g_{0L}^2+g_{1L}^2-g_{1R}^2}\right)\right]\ .
\end{eqnarray}
Expanding the gauge eigenstates in terms of the mass eigenstates, to ${\cal O}(a)$, we obtain
\begin{eqnarray}
W_{0L}^{\pm\mu}&\approx&W^{\pm\mu}\left(\cos\phi_L+\frac{a}{4}\sin4\phi_L\sin{\phi_L}\right)
+W_L^{\pm\mu}\left(-\sin\phi_L+\frac{a}{4}\sin{4\phi_L}\cos\phi_L\right)\nonumber\\
&&+W_R^{\pm\mu}\left(-a\frac{g_L}{g_{1R}}\cos\phi_L+a\frac{g_{0L}g_{1R}\sin^2\phi_L}{
g_{0L}^2+g_{1L}^2-g_{1R}^2}\right)
\nonumber\\
W_{1L}^{\pm\mu}&\approx&W^{\pm\mu}\left(\sin\phi_L-\frac{a}{4}\sin{4\phi_L}\cos\phi_L\right)
+W_L^{\pm\mu}\left(\cos\phi_L+\frac{a}{4}\sin4\phi_L\sin{\phi_L}\right)\nonumber\\
&&+W_R^{\pm\mu}\left(-a\frac{g_L}{g_{1R}}\sin\phi_L-a\frac{g_{0L}g_{1R}\sin\phi_L\cos\phi_L}{
g_{0L}^2+g_{1L}^2-g_{1R}^2}\right)
\label{eq:ccmixing}\\
W_{1R}^{\pm\mu}&\approx&W^{\pm\mu}\left(a\frac{g_L}{g_{1R}}\right)
+W_L^{\pm\mu}\left(a\frac{g_{0L}g_{1R}\sin\phi_L}{
g_{0L}^2+g_{1L}^2-g_{1R}^2}\right)+W_R^{\pm\mu}
\ .\nonumber
\end{eqnarray}

%%%%%%%%%%%%%%%%%%%%%%%%%%%%%%%%%%%%%%%%%%%%%%%%%%%%%%%%%%%%%%%%%%%%%%%%%%%%%%%%%%%%%%
\subsection{The Neutral Sector}
\label{subsec:neutral-mass-estates}

The mass matrix for the neutral gauge fields takes the form:
\begin{equation}
{\cal{M}}_{NC}^2\, = \, 
\frac{f^2}{2}
\left(
\begin{array}{cccc} 
g_{{0L}}^2          		& -(1-a)g_{{0L}} g_{{1L}} 	& -a g_{{0L}} g_{{1R}}	& 0 			\\
-(1-a)g_{{0L}} g_{{1L}} 	&  g_{{1L}}^2  		& 0				& -a g_{{1L}} g_{{0R}}\\
-a g_{{0L}} g_{{1R}} 		& 0 				& g_{{1R}}^2 			& -(1-a)g_{{1R}} g_{{0R}} \\
0				& -a g_{{1L}} g_{{0R}}	& -(1-a)g_{{1R}} g_{{0R}} 	&  g_{{0R}}^2
\end{array}
\right) \,.\label{eq:NCmatrix}
\end{equation}
For $a=0$ the mass matrix is block diagonal, so that the $SU(2)_{0L}\times SU(2)_{1L}$ and the $SU(2)_{0R}\times SU(2)_{1R}$ sub-matrices can be diagonalized separately in terms of the angles $\phi_L$, defined in Eq.~(\ref{eq:phiL}), and $\phi_R$, defined similarly by
\begin{eqnarray}
		 \sin\phi_R & = & \frac{g_{{0R}}}{\sqrt{g_{{0R}}^2+g_{{1R}}^2}} \,,\\
		 \cos\phi_R & = & \frac{g_{{1R}}}{\sqrt{g_{{0R}}^2+g_{{1R}}^2}} \,. \label{eq:phiR}
\end{eqnarray}
The angle $\phi_R$ is related to the coupling $g_R$, from Eq.~(\ref{eq:gLgR}), by
\begin{eqnarray}
		 g_R & = & g_{{0R}}\cos\phi_R = g_{{1R}}\sin\phi_R \,.
\end{eqnarray}
After diagonalizing the two sub-matrices, there are two massless neutral states, which correspond to the
standard model $W^{3\mu}$ and $B^\mu$.  One linear combination of these is the photon, which is massless
for arbitrary values of the parameter $a$.  It can be separated out in terms of a third angle $\theta$ (essentially
the weak mixing angle), which is defined by
\begin{eqnarray}
		 \sin\theta & = & \frac{g_R}{\sqrt{g_L^2+g_R^2}} \,,\nonumber\\
		 \cos\theta & = & \frac{g_L}{\sqrt{g_L^2+g_R^2}} \,. \label{eq:weakmixing}
\end{eqnarray}
The coupling to the photon is
\begin{eqnarray}
		\frac{1}{e^2}&=&\frac{1}{g_{L}^2}+\frac{1}{g_{R}^2}\,=\,\frac{1}{g_{0L}^2}+\frac{1}{g_{1L}^2}+\frac{1}{g_{0R}^2}+\frac{1}{g_{1R}^2}\,,
\end{eqnarray}
so that $e=g_L\sin\theta=g_R\cos\theta$.

For nonzero vacuum expectation value, there will be four neutral states:  the photon $A^\mu$, which is exactly massless,
the light eigenstate $Z^\mu$, and two heavy eigenstates, $Z_L$ and $Z_R$.  We can solve perturbatively in the parameter $a$
for the masses and mixings of these states.  To ${\cal O}(a^2)$, the masses are
\begin{eqnarray}
M_A^2&=&0\qquad \mathrm{(exact)}\nonumber\\
M_Z^2&\approx&\frac{f^2}{2}\left[2a(g_L^2+g_R^2)-a^2(g_L^2+g_R^2)\left(\cos^2{2\phi_L}+\cos^2{2\phi_R}\right)\right]\nonumber\\
M_{Z_L}^2&\approx&\frac{f^2}{2}\left[(g_{0L}^2+g_{1L}^2)-2ag_L^2+a^2\left((g_L^2+g_R^2)\cos^2{2\phi_L}+\frac{G_{LR}^2}{\Delta g^2}\right)\right]\nonumber\\
M_{Z_R}^2&\approx&\frac{f^2}{2}\left[(g_{0R}^2+g_{1R}^2)-2ag_R^2+a^2\left((g_L^2+g_R^2)\cos^2{2\phi_R}-\frac{G_{LR}^2}{\Delta g^2}\right)\right]\ ,
\end{eqnarray}
where we have defined for compactness:
\begin{eqnarray}
G_{LR}&=&g_{0L}g_{1R}\sin\phi_L\cos\phi_R+g_{1L}g_{0R}\cos\phi_L\sin\phi_R\nonumber\\
\Delta g^2&=&g_{0L}^2+g_{1L}^2-g_{0R}^2-g_{1R}^2\,.
\end{eqnarray}

Expanding the gauge eigenstates in terms of the mass eigenstates, we obtain
\begin{eqnarray}
W_{0L}^{3\mu}&\approx&A^\mu\left(\sin\theta\cos\phi_L\right)+
Z^{\mu}\left(\cos\theta\cos\phi_L+a\frac{\sin4\phi_L\sin{\phi_L}}{4\cos\theta}\right)\nonumber\\
&&
+Z_L^{\mu}\left(-\sin\phi_L+\frac{a}{4}\sin{4\phi_L}\cos\phi_L\right)
+Z_R^{\mu}\left(-a\frac{\sin4\phi_R\cos\theta\cos\phi_L}{4\sin\theta}+a\frac{G_{LR}\sin\phi_L}{
\Delta g^2}\right)
\nonumber\\
W_{1L}^{3\mu}&\approx&A^\mu\left(\sin\theta\sin\phi_L\right)+
Z^{\mu}\left(\cos\theta\sin\phi_L-a\frac{\sin{4\phi_L}\cos\phi_L}{4\cos\theta}\right)\nonumber\\
&&
+Z_L^{\mu}\left(\cos\phi_L+\frac{a}{4}\sin4\phi_L\sin{\phi_L}\right)
+Z_R^{\mu}\left(-a\frac{\sin4\phi_R\cos\theta\sin\phi_L}{4\sin\theta}-a\frac{G_{LR}\cos\phi_L}{
\Delta g^2}\right)
\nonumber\\
W_{1R}^{3\mu}&\approx&A^\mu\left(\cos\theta\sin\phi_R\right)+
Z^{\mu}\left(-\sin\theta\sin\phi_R+a\frac{\sin{4\phi_R}\cos\phi_R}{4\sin\theta}\right)\label{eq:ncmixing}\\
&&
+Z_L^{\mu}\left(-a\frac{\sin4\phi_L\sin\theta\sin\phi_R}{4\cos\theta}+a\frac{G_{LR}\cos\phi_R}{\Delta g^2}\right)
+Z_R^{\mu}\left(\cos\phi_R +\frac{a}{4}\sin4\phi_R\sin\phi_R\right)
\nonumber\\
B_{0R}^{\mu}&\approx&A^\mu\left(\cos\theta\cos\phi_R\right)+
Z^{\mu}\left(-\sin\theta\cos\phi_R-a\frac{\sin4\phi_R\sin{\phi_R}}{4\sin\theta}\right)\nonumber\\
&&
+Z_L^{\mu}\left(-a\frac{\sin4\phi_L\sin\theta\cos\phi_R}{4\cos\theta}-a\frac{G_{LR}\sin\phi_R}{\Delta g^2}\right)
+Z_R^{\mu}\left(-\sin\phi_R +\frac{a}{4}\sin4\phi_R\cos\phi_R\right)
\nonumber\,,
\end{eqnarray}
where the coefficients of $A^\mu$ are exact, while the other coefficients are correct to ${\cal O}(a)$.
 
%%%%%%%%%%%%%%%%%%%%%%%%%%%%%%%%%%%%%%%%%%%%%%%%%%%%%%%%%%%%%%%%%%%%%%%%%%%%%%%%%%%%%%
\section{Fermion Masses and Mixing in the Top Quark Sector}
\label{sec:fermionmass}
The mass terms for the fermions can be obtained from Eqs.~(\ref{eq:lagrangebulk}) and~(\ref{eq:lagrangebrane}).
We are assuming that $\lambda_3=0$, and that $\lambda_1$ and $\lambda_2$ are small for all fermions, except for the
top quark.  Thus, the only Yukawa coupling that is non-negligible is $\lambda_{1}$ for the top quark sector, and the only fermions for which there will be substantial mixing are in the top quark sector.  In addition, 
this Yukawa term only mixes charge $+2/3$ quarks, so that we need only be concerned with them.  

There are nine charge $+2/3$ quarks of each chirality in the top quark sector.  Their mass terms in the Lagrangian are
\begin{eqnarray}
{\cal L}_{\rm top\ sector}&=& -\lambda_Af\left(\bar{\chi}_{L}^{tA}\chi_{R}^{tA}+\bar{t}_{L}^{A}t_{R}^{A}\right) -\lambda_Bf\left(\bar{Q}_{L}^{tB}Q_{R}^{tB}+\bar{\chi}_{L}^{tB}\chi_{R}^{tB}\right)\nonumber\\
&&
-\lambda_Cf\left(\bar{Q}_{L}^{tC}Q_{R}^{tC}+\bar{\chi}_{L}^{tC}\chi_{R}^{tC}
+\bar{\phi}_{L}^{tC}\phi_{R}^{tC}+\bar{t}_{L}^{C}t_{R}^{C}\right)\\
&&-\lambda_{1}f\left(\bar{t}_{L}^{A}c+\frac{is}{\sqrt{2}}\left(\bar{Q}_{L}^{tA}+\chi_{L}^{tA}\right)\right)
\left(\bar{t}_{R}^{B}c-\frac{is}{\sqrt{2}}\left(\bar{Q}_{R}^{tB}+\chi_{R}^{tB}\right)\right)
+{\rm h. c.} \ ,\nonumber
\label{eq:lagrangetop}
\end{eqnarray}
where $s=\sin(\sqrt{2}|H|/f)$ and $c=\cos(\sqrt{2}|H|/f)$.
The fields that come from the $\psi^C$ multiplets are not mixed by the $\lambda_{1}$ Yukawa-term.  They
combine to form four Dirac states with masses $M_C=\lambda_Cf$.  In addition, we can diagonalize one linear
combination of each of the $\psi^A$ and $\psi^B$ fields that do not appear in the $\lambda_{1}$ Yukawa-term.
Introducing the new linear combinations,
\begin{eqnarray}
Q^{tB}&=&\frac{1}{\sqrt{2}}\left(T^B+K^{tB}\right)\nonumber\\
\chi^{tB}&=&\frac{1}{\sqrt{2}}\left(T^B-K^{tB}\right)\nonumber\\
t^{A}&=&\frac{1}{\sqrt{1-s^2/2}}\left(cT^A+\frac{is}{\sqrt{2}}K^{tA}\right)\\
\chi^{tA}&=&\frac{1}{\sqrt{1-s^2/2}}\left(cK^{tA}+\frac{is}{\sqrt{2}}T^{A}\right)\nonumber
 \ ,\nonumber
\label{eq:topmixAB}
\end{eqnarray}
we find that the Dirac field $K^{tA}=(K^{tA}_{L},K^{tA}_{R})$ decouples with mass  $M_A=\lambda_Af$, and the Dirac field
 $K^{tB}=(K^{tB}_{L},K^{tB}_{R})$ decouples with mass $M_B=\lambda_Bf$.  
 
 The remaining set of three left-handed
 and right-handed fermions mix with a mass lagrangian given by
\begin{equation}
{\cal{L}_{\rm top\ mass}} = -\bar{\cal T}_L {\cal M}_{\rm top} {\cal T}_R\,+\,{\rm h. c.}\,,
\label{eq:Ltopmass}
\end{equation}
where
\begin{eqnarray}
{\cal T}_L = \left(
\begin{array}{c}
T^{A}_{L}\\
T^B_{L} \\ 
Q^{tA}_{L}
\end{array}
\right) \,,    \quad \quad {\cal T}_R = \left(
\begin{array}{c}
T^A_{R} \\
T^B_{R}  \\
t^{B}_{R}
\end{array}
\right)\,,
\end{eqnarray}
and
\begin{eqnarray}
{\cal{M}}_{\rm top} & = & 
f
\left(
\begin{array}{ccc} 
\lambda_A &\ -i\lambda_{1}s\sqrt{1-\frac{s^2}{2}} \  &\ \lambda_{1}c\sqrt{1-\frac{s^2}{2}}  \\
0 &\lambda_B & 0  \\
0 &\lambda_{1}\frac{s^2}{\sqrt{2}} &i\lambda_{1}\frac{sc}{\sqrt{2}} 
\end{array}
\right) \,.\label{eq:topmatrix}
\end{eqnarray}

This fermion mass matrix can be diagonalized with a biunitary transformation, $V{\cal M}U^\dagger$.
To simplify the following expressions, we recall the definition for the top Yukawa coupling, Eq.~(\ref{eq:lambdatop}), 
\begin{equation}
\lambda_t^2\,=\,\frac{\lambda_A^2\lambda_1^2}{\lambda_A^2+\lambda_1^2}\,.
\end{equation}
We also define
\begin{equation}
\Delta\lambda^2\ =\ \lambda_A^2+\lambda_1^2-\lambda_B^2\ .
\end{equation}
Then, to ${\cal O}(s^2)$, we obtain the mass of the light eigenstate (the top quark):
\begin{eqnarray}
 	m_t^2 = \frac{\lambda_t^2  f^2}{2}s^2 + \left[\frac{\lambda_t^6 f^2}{4\lambda^2_A\lambda_1^2} 
	-\frac{\lambda_t^4 f^2}{2\lambda_1^2}\right]s^4\,,
\end{eqnarray}
and the masses of the heavy eigenstates:
\begin{eqnarray}
 	m_{T^{A\prime}} &=& (\lambda^2_A+\lambda_1^2)f^2 + \left[ -\frac{\lambda_t^2  f^2}{2}  + \frac{\lambda_B^2 \lambda_1^2 f^2}{\Delta\lambda^2 } \right] s^2 \nonumber\\
	&&+\, \left[-\frac{\lambda_t^6  f^2}{4\lambda^2_A\lambda_1^2} 
	+\frac{\lambda_t^4  f^2}{2\lambda_1^2} - \frac{\lambda_B^4 \lambda_1^4 f^2}{(\Delta\lambda^2)^3} -  \frac{\lambda_A^2 \lambda_1^2 (\lambda^2_A -\lambda_B^2) f^2}{2(\Delta\lambda^2)^2}\right]s^4
\end{eqnarray}
and
\begin{eqnarray}
 	m_{T^{B\prime}}  &=& \lambda^2_B f^2 + \left[ - \frac{\lambda_B^2 \lambda_1^2 f^2}{\Delta\lambda^2} \right] s^2 + \left[\frac{\lambda_B^4 \lambda_1^4 f^2}{(\Delta\lambda^2)^3} +  \frac{\lambda_A^2 \lambda_1^2 (\lambda^2_A -\lambda_B^2) f^2}{2(\Delta\lambda^2)^2}\right]s^4\,.
\end{eqnarray}
 
To ${\cal O}(s^2)$, the left-handed gauge eigenstates in terms of mass eigenstates are
\begin{eqnarray}
 	Q^{tA}_{L} &=& \left(1-\frac{s^2}{4} \frac{\lambda_t^4}{\lambda_A^4}\right) t_L + \frac{is}{\sqrt{2}}\frac{\lambda_t^2}{ \lambda_A^2} T^{A\prime}_L + \frac{s^2}{\sqrt{2}}\frac{\lambda_1 }{\lambda_B}\frac{\lambda_A^2-\lambda_B^2}{\Delta\lambda^2} T^{B\prime}_L  \,,   \\
	T^A_{L} &=& \left(1-\frac{s^2}{4} \frac{\lambda_t^4}{\lambda_A^4}-\frac{s^2}{2} \frac{\lambda_1^2\lambda_B^2}{(\Delta\lambda^2 )^2}\right) T^{A\prime}_L +  \frac{i s}{\sqrt{2}}\frac{\lambda_t^2}{ \lambda_A^2} t_L + is\frac{\lambda_1 \lambda_B}{\Delta\lambda^2 } T^{B\prime}_L\,,\\
	T^B_{L}  &=& \left(1-\frac{s^2}{2} \frac{\lambda_1^2\lambda_B^2}{(\Delta\lambda^2)^2}\right) T^{B\prime}_L - \frac{s^2}{\sqrt{2}}\frac{ \lambda_t^2}{\lambda_B \lambda_1} t_L + is\frac{\lambda_1 \lambda_B}{\Delta\lambda^2 } T^{A\prime}_L \,,
\end{eqnarray}
while the right-handed gauge eigenstates in terms of mass eigenstates are
\begin{eqnarray}
t_R^B &=&-i \frac{\lambda_t}{\lambda_1}\left(1+\frac{s^2}{4} \frac{\lambda_1^2(\lambda_1^2 + 3\lambda_A^2)}{(\lambda_1^2 + \lambda_A^2)^2}\right) t_R + is\frac{\lambda_1^2 }{\Delta\lambda^2} T^{B\prime}_R\nonumber\\
	&&+\, \frac{\lambda_t}{ \lambda_A}\left(1-\frac{s^2}{2} \frac{\lambda_1^2(\lambda_1^2 + \lambda_A^2)}{(\Delta\lambda^2 )^2}-\frac{s^2}{4} \frac{\lambda_A^2(\lambda_1^2 + 3\lambda_A^2)}{(\lambda_1^2 + \lambda_A^2)^2}\right)T^{A\prime}_R \,,
 	\\
	T^A_{R} &=& \frac{\lambda_t}{\lambda_1}\left(1-\frac{s^2}{2} \frac{\lambda_1^2(\lambda_1^2 + \lambda_A^2)}{(\Delta\lambda^2 )^2}+\frac{s^2}{4} \frac{\lambda_1^2(\lambda_1^2 + 3\lambda_A^2)}{(\lambda_1^2 + \lambda_A^2)^2}\right) T^{A\prime}_R\nonumber\\
	&&+i\, \frac{\lambda_t}{ \lambda_A}\left(1-\frac{s^2}{4} \frac{\lambda_A^2(\lambda_1^2 + 3\lambda_A^2)}{(\lambda_1^2 + \lambda_A^2)^2}\right) t_R
	+  is\frac{\lambda_1 \lambda_A}{\Delta\lambda^2} T^{B\prime}_R\,,\\
	T^B_{R} &=& \left(1- \frac{s^2}{2}\frac{\lambda_1^2(\lambda_1^2 + \lambda_A^2)}{(\Delta\lambda^2 )^2}\right) T^{B\prime}_R  + is\frac{\lambda_t (\lambda_1^2 + \lambda_A^2) }{\lambda_A(\Delta\lambda^2) } T^{A\prime}_R\,.
\end{eqnarray}

%%%%%%%%%%%%%%%%%%%%%%%%%%%%%%%%%%%%%%%%%%%%%%%%%%%%%%

\section{Higgs Potential for small $|H|/f$}
\label{sec:Hpot}

At small values of the Higgs field $H$, the one-loop Coleman-Weinberg potential can be 
expanded as
\begin{equation}
V\ =\ m^2H^\dagger H\,+\,\lambda(H^\dagger H)^2+\cdots\,,
\end{equation}
where the coupling $\lambda$ will also have logarithmic dependence on $H^\dagger H$.
Letting $m^2=m^2_{\rm gauge}+m^2_{\rm fermion}$, we have
\begin{equation}
m^2_{\rm gauge}\ =\ \frac{3}{64\pi^2}\left\{3M_{W_L}^2g_L^2\left(\ln\frac{\Lambda^2}{M_{W_L}^2}-\frac{1}{2}\right) +M_{Z_R}^2g_R^2\left(\ln\frac{\Lambda^2}{M_{Z_R}^2}-\frac{1}{2}\right) \right\}\ ,
\end{equation}
with $M_{W_L}^2=M_{Z_L}^2=(g_{0L}^2+g_{1L}^2)f^2/2$,  $M_{W_R}^2=g_{1R}^2f^2/2$ and $M_{Z_R}^2=(g_{0R}^2+g_{1R}^2)f^2/2$,
and
\begin{equation}
m^2_{\rm fermion}\ =\ \frac{3}{8\pi^2}\left\{\left(2M_{T_B}^2\lambda_1^2-M_{T_A}^2\lambda_t^2\right)
\left(\ln\frac{\Lambda^2}{M_{T_A}^2}-\frac{1}{2}\right) +\frac{2M_{T_B}^4\lambda_1^2}
{M_{T_A}^2-M_{T_B}^2}\ln\frac{M_{T_B}^2}{M_{T_A}^2}
 \right\}\ ,
\end{equation}
with $M_{T_A}^2=(\lambda_A^2+\lambda_1^2)f^2$ and $M_{T_B}^2=\lambda_B^2f^2$.

Expressing the $(H^\dagger H)^2$ coupling as $\lambda=\lambda_{\rm gauge}+
\lambda_{\rm fermion}$, we have
\begin{eqnarray}
\lambda_{\rm gauge}&=&-\frac{3}{256\pi^2}\left\{g_{0L}^2\left(g_{1L}^2+g_{1R}^2\right)\left(\ln\frac{\Lambda^2}{M_{W_L}^2}+\frac{M_{W_R}^2}{M_{W_R}^2-M_{W_L}^2}\ln\frac{M_{W_L}^2}{M_{W_R}^2}-\frac{1}{2}\right) \right.\nonumber\\
&&\qquad\qquad\ 
+2g_L^4\left(\ln\frac{M_{W_L}^2}{M_W^2(H)}-\frac{1}{2}\right) 
+\left[\frac{4g_L^2M_{W_L}^2M_{W_R}^2/f^2}{M_{W_L}^2-M_{W_R}^2}\right]\ln\frac{M_{W_L}^2}{M_{W_R}^2}\nonumber\\
&&\qquad\qquad\ 
+{\half}(g_{0L}^2+g_{0R}^2)(g_{1L}^2+g_{1R}^2)\left(\ln\frac{\Lambda^2}{M_{Z_L}^2}
+\frac{M_{Z_R}^2}{M_{Z_R}^2-M_{Z_L}^2}\ln\frac{M_{Z_L}^2}{M_{Z_R}^2}-\frac{1}{2}\right)\nonumber\\
&&\qquad\qquad\ 
+g_{L}^4\left(\ln\frac{M_{Z_L}^2}{M_Z^2(H)}-\frac{1}{2}\right)
+g_{R}^4\left(\ln\frac{M_{Z_R}^2}{M_Z^2(H)}-\frac{1}{2}\right)\\
&&\qquad\qquad\ 
+2g_{L}^2g_R^2\left(\ln\frac{M_{Z_L}^2}{M_Z^2(H)}
+\frac{M_{Z_L}^2}{M_{Z_L}^2-M_{Z_R}^2}\ln\frac{M_{Z_R}^2}{M_{Z_L}^2}
+\frac{1}{2}\right)\nonumber\\
&&\qquad\qquad\ \left.
+\left[\frac{2(g_L^2+g_R^2)M_{Z_L}^2M_{Z_R}^2/f^2}{M_{Z_L}^2-M_{Z_R}^2}\right]\ln\frac{M_{Z_L}^2}{M_{Z_R}^2}\right\}\ \ -\ \frac{m^2_{\rm gauge}}{6f^2}\nonumber\ .
\end{eqnarray}
and
\begin{eqnarray}
\lambda_{\rm fermion}&=&\frac{3}{4\pi^2}\left\{\frac{\lambda_t^4}{4}\left(\ln\frac{M_{T_A}^2}{M_t^2(H)}-\frac{1}{2}\right)-\ln(1-x)\left[\frac{\lambda_1^4(2-x)}{x^3}+\frac{\lambda_1^2\lambda_t^2(1-x)}{x^2}+\frac{\lambda_1^2\lambda_A^2}{x}\right]\right.\nonumber\label{eq:lambdafermion}\\
&&\qquad\quad \left.-\left[\frac{2\lambda_1^4}{x^2}+\frac{\lambda_1^2\lambda_t^2}{x}\right]
\right\}-\frac{\ \ 2m^2_{\rm fermion}}{3f^2}\ ,
\end{eqnarray}
where $x=1-M_{T_A}^2/M_{T_B}^2$.  In addition, in the above formulae, we use the field-dependent masses for the light fields: $M_W^2(H)=g_L^2(H^\dagger H)/2$, $M_Z^2(H)=(g_L^2+g_R^2)(H^\dagger H)/2$, and $M_t^2(H)=\lambda_t^2(H^\dagger H)$.

%%%%%%%%%%%%%%%%%%%%%%%%%%%%%%%%%%%%%%%%%%%%%%%%%%%%%%

\section{Fermion Sector with Complete $SO(5)$ Multiplets and Decoupled SM Partners}

\label{sec:fermionmod}

In order to probe the sensitivity of the model to UV completion of the fermion sector, we consider a modification that
leaves the fermion contribution to the effective potential completely finite at one loop\footnote{We are grateful to an anonymous referee for suggesting this modification of the fermion sector.}.
First, we make the fields $\psi^A_R$ and $\psi^B_L$ into complete $SO(5)$ multiplets by
reinstating the missing SM partners, $Q^A_R$ and $u^B_L$, in Eqs.~(\ref{eq:psia}) and (\ref{eq:psib}).  Then we decouple them
by adding two new fermions, $Q^{\prime A}_L$ and $u^{\prime B}_R$, which mix via large mass terms,\begin{eqnarray}
\Delta{\cal L}_{\rm mass}&=& -\Lambda^\prime_A\bar{Q}^{\prime A}_LQ^A_R-\Lambda^\prime_B\bar{u}^{ B}_Lu^{\prime B}_R+{\rm h. c.} 
\ .
\label{eq:lagrangedecouple}
\end{eqnarray}
With this modification, the Dirac mass terms proportional to $\lambda_A$ 
and $\lambda_B$ of Eq.~(\ref{eq:lagrangebulk}) now preserve both the $SO(5)_0$ and $SO(5)_1$ symmetries, since the
Dirac fields $\psi^A$ and $\psi^B$ are in complete $SO(5)$ multiplets.  Instead, the collective symmetry breaking occurs 
through the Yukawa terms of Eq.~(\ref{eq:lagrangebrane}), which break the $SO(5)_1$ symmetry, and the decoupling mass terms of Eq.~(\ref{eq:lagrangedecouple}), which break the $SO(5)_0$ symmetry.  However, these two symmetry-breaking terms contain no fermion fields in common; therefore, any one-loop diagram that contributes to the Higgs potential and breaks both $SO(5)$ symmetries must contain Dirac mass insertions to mix the fermion fields (in addition to the two symmetry-breaking insertions).  The requirement of the three separate contributions to the one-loop diagrams renders them completely finite.  

With the modified fermion sector, the masses of all of the original eigenstates are unchanged, up to corrections of ${\cal O}(f^2/\Lambda^{\prime 2}_{A,B})$.  In addition, there are two new heavy eigenstates with Higgs-field-dependent masses given by
\begin{eqnarray}
 	M^2_{\Lambda_{A}} &=& \Lambda^{\prime 2}_A+\lambda^2_Af^2+\frac{\lambda_1^2\lambda_A^2f^4}{\Lambda^{\prime 2}_A}\frac{s^2}{2}+\cdots\nonumber\\
	M^2_{\Lambda_{B}} &=& \Lambda^{\prime 2}_B+\lambda^2_Bf^2+\frac{\lambda_1^2\lambda_B^2f^4}{\Lambda^{\prime 2}_B}c^2+\cdots\ .
\end{eqnarray}
where $s=\sin(\sqrt{2}|H|/f)$ and $c=\cos(\sqrt{2}|H|/f)$, and we have neglected terms of ${\cal O}(f^6/\Lambda^{\prime 4}_{A,B})$.  Including the effects of the heavy mass eigenstates in the 
Coleman-Weinberg effective potential gives a new contribution of
\begin{eqnarray}
\Delta V_{\rm fermion}&=&-\frac{3}{16 \pi^2}f^4\lambda_1^2s^2\left\{2\lambda_B^2
\left(\ln\frac{\Lambda^2}{\Lambda_{B}^{\prime 2}}-\frac{1}{2}\right) - \lambda_A^2
\left(\ln\frac{\Lambda^2}{\Lambda_{A}^{\prime 2}}-\frac{1}{2}\right)
 \right\}\ .
 \end{eqnarray}
 Redefining $\Lambda^\prime_{A,B}=e^{-1/4}\Lambda_{A,B}$, we obtain
\begin{eqnarray}
\Delta V_{\rm fermion}&=&-\frac{3}{16 \pi^2}f^4\lambda_1^2s^2\left\{2\lambda_B^2
\ln\frac{\Lambda^2}{\Lambda_{B}^{ 2}} - \lambda_A^2
\ln\frac{\Lambda^2}{\Lambda_{A}^{2}}
 \right\}\ ,\label{eq:Vmod}
 \end{eqnarray}
 which is exactly the modified potential studied in section \ref{sec:potential}.  As expected, the dependence on the
UV cutoff $\Lambda$ in Eq.~(\ref{eq:Vmod}) exactly cancels with that from the other fermion fields, exchanging it for a dependence on the scales
 $\Lambda_A$ and $\Lambda_B$.

%%%%%%%%%%%%%%%%%%%%%%%%%%%%%%%%%%%%%%%%%%%%%%%%%%%%%%

\end{document}